\pdfoutput=1
\documentclass[12pt]{article}
\usepackage[utf8]{inputenc}
\usepackage{empheq}
\usepackage{bm}
\usepackage[normalem]{ulem}
\usepackage{latexsym}
\usepackage{color}
\usepackage{amsmath,amsfonts,amssymb}
\usepackage{graphicx,epsfig}
\usepackage{psfrag}
\usepackage{amsthm}
\usepackage{cite,url}
\usepackage{makecell}
\usepackage{float}

\pdfoutput=1

\usepackage{verbatim}   % To comment out a full section by \begin{comment}...\end{comment}

\usepackage{amsmath}
\usepackage{amssymb}
\usepackage{amsfonts}
\usepackage{mathrsfs}

\usepackage{mathrsfs}
\usepackage{fullpage}
\usepackage{setspace}
\usepackage{bm}
\usepackage{bbm}

\usepackage[normalem]{ulem}
\usepackage{enumerate}
\usepackage{yfonts}

\usepackage{psfrag}

\usepackage{graphicx}
\usepackage{cancel}
\usepackage{slashed}

\usepackage[font=small,labelfont=bf]{caption}
\usepackage{subcaption}

\newcommand{\be}{\begin{equation}}
\newcommand{\bea}{\begin{eqnarray}}
\newcommand{\ee}{\end{equation}}
\newcommand{\eea}{\end{eqnarray}}

\begin{document}
\numberwithin{equation}{section}
{
\begin{titlepage}
\begin{center}

\hfill \\
\hfill \\
\vskip 0.2in

{\Large Gravitational radiation from hyperbolic encounters in the presence of dark matter}\\

\vskip .7in

{\large
Abhishek Chowdhuri \footnote{\url{chowdhuri_abhishek@iitgn.ac.in}}, Rishabh Kumar Singh \footnote{\url{singhrishabh@iitgn.ac.in}}, Kaushik Kangsabanik \footnote{\url{kaushikkangsabanik17@gmail.com}},\\ Arpan Bhattacharyya \footnote{\url{abhattacharyya@iitgn.ac.in}}
}

\vskip 0.3in

{\it Indian Institute of Technology, Gandhinagar, Gujarat-382355, India}\vskip .5mm

\vskip.5mm

\end{center}

\vskip 0.35in

\begin{center} {\bf ABSTRACT } \end{center}
In this study, we look into binaries undergoing gravitational radiation during a hyperbolic passage. Such hyperbolic events can be a credible source of gravitational waves in future detectors. We systematically calculate fluxes of gravitational radiation from such events in the presence of dark matter with different profiles, also considering the effects of dynamical friction. We provide an estimate for the braking index and show how it evolves due to the presence of the dark matter medium. We also investigate the binary dynamics through the changes in the orbital parameters by treating the potential due to dark matter spike and the dynamical friction effects as a perturbation term. An insight into the effects of such a medium on the binaries from the corresponding osculating elements opens up avenues to study binary dynamics for such events.
\vfill

%\noindent \today

\end{titlepage}
}

%%%%%%%%%%%%%%%%%%%%%%%%%%%%%%%%%%%%%%%%%%%%%%%%%%%%%%%%%%%%%%%%%%%%%%%%%%%%%%%%
% Table of contents
%%%%%%%%%%%%%%%%%%%%%%%%%%%%%%%%%%%%%%%%%%%%%%%%%%%%%%%%%%%%%%%%%%%%%%%%%%%%%%%%
\newpage
\tableofcontents
%%%%%%%%%%%%%%%%%%%%%%%%%%%%%%%%%%%%%%%%%%%%%%%%%
\section{Introduction}
General Relativity has repeatedly proved to be the most robust theory of gravity, passing various tests\cite{Hoyle:2000cv,2001CQGra..18.2397A,1993tegp.book.....W, Will:2014kxa, Stairs:2003eg, Wex:2014nva, Manchester:2015mda, Kramer:2016kwa}. With the advent of LIGO-VIRGO detectors, General Relativity is now being tested using the outpouring data we receive from consecutive runs\cite{Abbott:2016blz, TheLIGOScientific:2016wfe, Abbott:2016nmj, Abbott:2017vtc, LIGOScientific:2018mvr, LIGOScientific:2020ibl, LIGOScientific:2021djp, LIGOScientific:2017vwq, LIGOScientific:2017zic, LIGOScientific:2019hgc}. The detections are from inspiralling compact binaries and have brought a new era in gravitational wave (GW) precision astronomy. Compact binary scattering events can also be credible radiation sources in future detectors\cite{Kocsis:2006hq,OLeary:2008myb,Mukherjee:2020hnm}. These events can occur at very high eccentricities\cite{Hansen, Will}. Studies like the one shown in \cite{LIGOScientific:2014pky, OLeary:2008myb} have estimated the reasonable detection rates for such binary encounters to be around a few to thousands per year, comparable to that of inspiralling coalescing binaries. If one of the objects of these hyperbolic encounters turns out to be a neutron star, then we also have a possibility of electromagnetic signatures. These scenarios are astrophysically engaging as they can lead to a GW burst, and the estimated luminosity from such emissions is enormous\cite{Tsang:2013mca}.
\par
The flux calculation for gravitational waves from binaries in such hyperbolic orbits has been studied  using quadrupolar approximations in \cite{Turner, Capozziello:2008ra, DeVittori:2012da, Garcia-Bellido:2017knh}. Then it is extended upto 1PN order in \cite{Schaefer} and \cite{Damour} extended it further for quasi-Keplerian parametrizations. Studies in \cite{DeVittori:2014psa} took the calculations further to 1.5PN invoking in the same quasi-Keplerian parametrizations as done in \cite{Damour}. Taking true anomaly parametrizations \cite{Wex, Wex2} into account,  quadrupolar energy and angular momentum fluxes, as well as the 1PN amplitude corrected waveform have been studied in \cite{Majar:2010em, Majar:2008zz, Wagoner, Blanchet}. Similarly, a 1.5PN amplitude corrected waveform with Quasi-Keplerian parametrizations were also studied in\cite{DeVittori:2014psa}. Furthermore, a 3.5 PN accurate orbitals dynamics for non-spinning binaries on a hyperbolic track is given in \cite{Cho:2018upo}. In \cite{DeVittori:2012da}, authors gave a general analytical formula for the GW energy spectrum of compact binaries in unbound orbits generalizing the computations done in the parabolic limit by \cite{Berry:2010gt} and in \cite{Capozziello:2008ra}, authors gave an estimation of the expected number such close gravitational flybys towards different targets. However, studies like the one in \cite{Rubbo:2006dv} suggests that LISA could be better suited for the detection of GW burst signals associated with stellar mass compact objects in unbound orbits around massive black hole (BH)s. Event rates for such bursts and the associated GW measurements have been reported in \cite{Berry:2012im, Berry:2013poa, Berry:2013ara}.
\par 
Our universe is not empty but filled with non-trivial matter medium. Dark Matter (DM) is the most ubiquitous, whose nature has yet to be fully known, and the physics is still speculative. One of the most sought-after avenues deals with modelling these dark matter particles as \textit{WIMPs} (Weakly Interacting Massive Particles) and exploring their detection prospects both directly and indirectly\cite{Jungman:1995df, Bergstrom:2000pn, Boehm:2002yz, Boehm:2003bt, Serpico:2004nm, Steigman:2013yua, Nollett:2013pwa, Escudero:2018mvt, Sabti:2019mhn, Griest:1989wd}. The density profile of such WIMPs is universal, with a cusp near the galactic centre and also forming a spike near the central massive black hole. In recent times the advent of data from the LIGO-VIRGO detectors has been much interest in probing DM with GW \cite{Bhattacharya:2023stq,Singh:2022wvw, Baryakhtar:2022hbu}. Studies in \cite{Eda:2013gg} also supported the fact that phases of GWs can be modified, and such modifications can impact future space-borne GW experiments. In \cite{Macedo:2013qea,Barausse:2014pra}, dynamical friction effects from such dark matter spikes showed changes to the phase of a GW from binaries with circular orbits. 
\par
Motivated by these, in this paper, we study flyby events in presence of certain type of dark-matter halo. The orbit is specified to be hyperbolic one and the parametrization of it is true anomaly type \cite{Wex, Wex2}. Apart from computing GW fluxes emitted from such binaries, in this paper we mainly focus on studying the binary dynamics by taking into account the effects of dark matter, the dynamical friction provided by the medium in which these binaries are moving, the accretion effects and possible GW backreaction on the binaries. The dynamical friction and the accretion effects contribute to the conservative sector of the binary dynamics. We make a comparative study to see which effects are more predominant than the other by taking the Keplerian perturbation theory route. We model the standard Kepler problem, which admits hyperbolic orbits as one of the solutions for a given particle's total energy, and then use the method of the standard osculating elements to calculate the rate of change of orbital parameters. These equations come in handy to again investigate the binary dynamics in such encounters, giving us a scope for a comparative study with elliptical orbit cases as well, while also giving room for one to investigate into the waveform modelling aspect of the binaries. The work is relevant since a possible waveform model from such analysis can serve as a good prospect for the future detectors to observe such events and also put some insights into the parameters of the theory from GW data  coming from LIGO-Virgo-KAGRA and the International Pulsar Timing Array (PTA) along the lines of \cite{Dandapat:2023zzn}. 
\par
We organize the paper as follows: In Sec.~(\ref{sec1}), we briefly outline the basic setup for such binaries moving in hyperbolic tracks and compute the radiation fluxes from such systems. In Sec.~(\ref{sec2}), we briefly discuss the particular model of the DM distribution that we will consider throughout this paper. In Sec.~(\ref{sec3}), we compute GW fluxes due to hyperbolic encounters by taking into effect of the surrounding DM medium. In Sec.~(\ref{barkingindex}), we provide an estimate of the braking index, which may provide a possible observational signature of dark matter (through its dissipative effects) via GW wave signal emitted from hyperbolic encounters. Finally, in Sec.~(\ref{sec4}), we include the effect of this DM medium (as well as the GW backreaction) as a perturbative term in the Keplerian orbit equations and calculate the changes in orbital parameters due to these extra effects. We also discuss the details of the numerical method we used to solve the simultaneous equations and list our conclusions therein. Finally, we summarize our results and conclude with future directions with Sec.~(\ref{sec9}).
\par
\textit{Notation and Conventions}: We use units where the Gravitational constant and Planck constant is set to unity: $\textit{G}=\hbar=1.$ Also, the solar mass value $M_{\odot}$, that we have used all over the text is taken to be $10^{30}$ kg.

%%%%%%%%%%%%%%%%%%%%%%%%%%%%%%%%%%%%%%%%%%%%%%%%%

 \section {Binaries in hyperbolic orbits}\label{sec1}

We begin this section by briefly reviewing the basics of binary dynamics for hyperbolic orbits. The problem we are dealing with here is a typical two-body problem with masses $m_{1}$ and $m_{2}$ with the mass $m_{2}$ at the centre of force of the trajectory and the mass $m_{1}$ undergoing the scattering track around $m_{2}$. The usual way to approach such a problem is to reduce the problem to an effective one-body problem with a reduced mass $\mu=\frac{m_{1}m_{2}}{m_{1}+m_{2}}$ and total mass $M=m_{1}+m_{2}$, such that the total energy and the angular momentum of the system becomes 
\begin{align}
\begin{split} \label{eq:2.1}
 &      E=T+V=\frac{1}{2}\mu(\dot{r}^{2}+r^{2}\dot{\phi}^{2})+V(r)=const\,, \\& L_{z}=\mu r^{2}\dot{\phi}=const\,,
\end{split}
\end{align}
where the equations are written in plane-polar coordinates $(r,\phi)$ and %the coordinates are in the orbital plane of the trajectory with 
the potential being 
\begin{align}
    V(r)=-\frac{\mu M}{r}\,.
\end{align}
The solutions for such radial equations of motion are the well-known conic sections, with $r$ being parametrized by $\phi$:
\begin{equation} \label{eq:2.4}
    r(\phi)=\frac{a(e^{2}-1)}{1+e\cos(\phi-\phi_{0})}\,,
\end{equation}
with $$a=\frac{\mu M}{2E}\,\quad \textrm{and}\, e=\sqrt{1+\frac{2EL^{2}}{M^{2}\mu^{3}}}\,.$$ The semi-major axis, defined above and correspondingly the eccentricity $e$, can be expressed in terms of the binary parameters: $v_{0}, m_{1}, m_{2},b$. This can be done since for hyperbolic orbits $E$ and $L$ are related to the impact parameter $b$ and initial velocity $v_{0}$ as $$L=\mu v_{0}b\,\quad \textrm{and}\quad E=\frac{1}{2}\mu v^{2}_{0}\,.$$ Thus, the eccentricity becomes $e=\sqrt{1+\frac{v^{4}_{0}b^{2}}{M^{2}}}$ and $a=\frac{M}{v^{2}_{0}}$. One can also relate $\phi_{0}$ which is the angle subtended at the minimal distance $r_{0}$, i.e., the radius at the periastron, to eccentricity through the following relation: \begin{equation} e=-\frac{1}{\cos\phi_{0}}\,.\end{equation} 

\begin{figure} [htb!]
    \centering
    \includegraphics[ scale = .9]{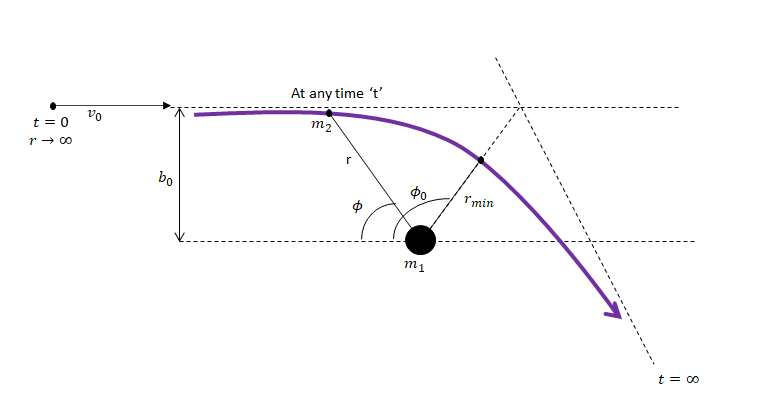}
    \caption{Geometry of hyperbolic orbit.}
    \label{gEOM}
\end{figure}

Considering these, the above conic section form can be written exclusively for a hyperbolic track as shown in Fig.~(\ref{gEOM}):
\begin{equation}\label{2.4}
    r(\phi)=\frac{b\sin\phi_{0}}{\cos(\phi-\phi_{0})-\cos\phi_{0}}\,.
\end{equation}
\par

This form can give the velocity of the masses moving in such trajectories in terms of $\phi_{0}$
\begin{equation}
    \label{eq:2.7}
    v= \sqrt{\dot r^2 + r^2 \dot \phi^2}=\frac{v_o}{\sin(\phi_{0})} \sqrt{1 + \cos^2(\phi_{0}) - 2 \cos(\phi_{0})\cos(\phi - \phi_{0})}\,.
\end{equation}
Given that one of the objects in the binary system is moving in such a trajectory around the central object, we can calculate the radiation emanated by using the orbit equation specific to the case of hyperbolas and then use the usual quadrupole formula to calculate the energy radiated via the GW. We start with the usual formula for this radiated power in the quadrupolar order, which is the lowest order approximation and is given by 
\cite{Maggiore}\,,
\begin{equation}
    \label{eq:3.1}
    P_{quad}=\frac{1}{45c^{5}}\langle \dddot{D}_{ij}\dddot{D}_{ij}\rangle \,.
\end{equation}
where $D_{ij}=3M_{ij}-\delta_{ij}M_{kk}$ and $M_{ij}=\frac{1}{c^{2}}\int T^{00}x_{i}x_{j}d^{3}x$, with the $00$-th component of the energy-momentum tensor being given by $T^{00}=\mu \delta(\vec{x}-\vec{x}^{\prime})c^{2}$. In terms of $M_{ij}$'s the expression for $P_{quad}$ becomes 
\begin{equation}
    \label{eq:3.2}
    P_{quad}=\frac{1}{45c^{5}}6\langle \dddot{M}^{2}_{11}+\dddot{M}^{2}_{22}+3\dddot{M}^{2}_{12}-\dddot{M}^{2}_{11}\dddot{M}^{2}_{22}\rangle\,.
\end{equation}
Then using (\ref{2.4}) we can get the expression for the radiated power which is\cite{Garcia-Bellido:2017knh}\,,
\begin{align}
    \begin{split}
        \label{eq:4.03}
        &  P_{quad}=-\frac{32L^{6}}{45c^{5}b^{8}\mu^{4}}f(\phi,\phi_{0}) \\
        &  f(\phi, \phi_{0})=  \frac{\sin^4(\frac{\phi}{2}) \sin^4(\frac{\phi}{2} - \phi_{0})}{\tan^2(\phi_{0}) \sin^6(\phi_{0})} \left [ 150 + 72 \cos(2\phi_{0})+66 \cos(2\phi_{0} - 2\phi) -144(\cos(\phi- 2\phi_{0})+ \cos(\phi))\right] \,.
        \end{split}
\end{align}

Note that the above expression for power radiation is purely due to the GW emission. It does not include the effect of the environment, which we consider here to be dark matter providing dynamical friction \cite{Chandra}, and the scenario has all the orbital parameters fixed while undergoing the dynamics. If one includes the effects of the medium in these radiative calculations, then additional terms will appear with the medium's details in them. Also, if we expect our radiation field to backreact on our system, the usual scenarios may be non-trivial. We explore these curiosities in the following sections, starting with calculating fluxes for such binaries and doing a comparative study on their relative magnitudes.

%%%%%%%%%%%%%%%%%%%%%%%%%%%%%%%%%%%%%%%%%%%%%%%%%%%%%%%%%%%%%%%%%%%%%%%%%%%%%%%%%%%%%%%%%%%%%%%%%%%%%%%%%%%%%%%%%%%%%%%%%%%%%%%%%%%%%%%%%%%%%%%%%%%%%%%%%%%%%%%%%%%%%%%%%%%%%%%%%%%%%%%%%%%%%%%%%%%%%%%%%

\section{Model for the dynamical friction  due to DM spike}\label{sec2}

We begin this section by exploring the dynamics of binaries moving in a hyperbolic orbit in the presence of a medium, more precisely focusing on the fluxes of radiation. A natural question to ask is about the nature of the medium. Around $27\%$ of the universe's total mass seems to come from the dark matter, and there are observational evidence in favor of it\cite{Bertone:2004pz, Clowe:2006eq, Planck:2018vyg, Bertone:2018krk, XENON:2018voc, Clark:2020mna, XENON:2023ysy, Tovey:2000mm, Freytsis:2010ne, Hardy:2015boa, DarkSide-20k:2017zyg, DARWIN:2016hyl, IceCube:2021kuw, DEAPCollaboration:2021raj, DEAP:2019yzn, DarkSide:2018kuk, LUX:2016ggv, PandaX-II:2017hlx, SuperCDMS:2017mbc, PICO:2019vsc, Bramante:2018qbc, Bhoonah:2018wmw, Bramante:2019yss, Huang:2018pbu}. They are ubiquitous, and more interestingly, Black Holes (BHs) in such DM environments build up halos around them due to the effect of the strong gravity. For our studies in this paper we will consider Navarro-Frenk-White (NFW) profile \cite{Navarro:1996gj} which form spikes due to the presence of certain overdense regions\cite{Gondolo:1999ef, Merritt:2002vj, Ullio:2001fb, Bertone:2005xv, Vasiliev:2008uz}. These overdense regions, in turn, result from the adiabatic growth of supermassive BHs with masses in the range $10^{6}\sim10^{9}M_{\odot}$ \cite{Gondolo:1999ef}. It seems logical that an object moving around a central BH might have its orbital motion affected by such spike structures. Optical observations of such activities can lead to indirect tests of the existence of such DM spikes and constrain the spike's density profile \cite{Bertone:2004pz, Fields:2014pia, Shelton:2015aqa, Takamori:2020ntj}. LIGO/VIRGO and other high-end detectors are gearing up to detect GWs from such binaries \cite{Coogan:2021uqv, Kavanagh:2020cfn, Eda:2013gg, Eda:2014kra, Yue:2017iwc, Yue:2018vtk, Hannuksela:2019vip, Cardoso:2019rou, Cardoso:2021wlq, Becker:2022wlo, Chan:2022gqd, Dai:2023cft, Cole:2022ucw, Dai:2021olt,Figueiredo:2023gas, Destounis:2022obl}, shedding light on DM features based on GW observations. 
\par 
 We consider a DM mini spike model given in \cite{Zhao:2005zr, Bertone:2005xz, Sadeghian:2013laa, Ferrer:2017xwm, Gondolo:1999ef, Quinlan:1994ed}. The density distribution is spherically symmetric with a power law behavior as shown below,
\begin{equation}
    \rho_{DM}=\begin{cases}
    \rho_{sp}\Big(\frac{r_{sp}}{r}\Big)^{\alpha} & ,\text{when } r_{min}\leq r\leq r_{sp}\\
        0 & ,\text{when } r \leq r_{min}
        \end{cases} \label{model}
\end{equation}
where $\rho_{sp}$ is a normalization constant, and $r_{sp}$ is the radius of the mini spike.
 There are estimates as to what this $\alpha$ value can be. Typically this range is: $2.25\leq \alpha \leq 2.5$ \cite{Gondolo:1999ef, Ullio:2001fb, Quinlan:1994ed, Young}. Henceforth we will strictly follow this range. For our subsequent analysis we take the the mass of the central massive object as $m_1= 10^{3}M_{\odot}$ and the mass of the secondary as $m_2= 10M_{\odot}.$ Also,  $\rho_{sp}=226 M_{\odot}/pc^{3}$ and $r_{sp}$ to be around $0.54\,pc$ following \cite{Eda:2013gg, Eda:2014kra, Dai:2021olt}.
\par
Such distributions lead us to give an equilibrium phase space distribution function, and in our case, the halo is spherically symmetric. The distribution function $f=f(\mathscr{E})$, where $\mathscr{E}$ is the relative energy per unit mass \cite{Becker:2021ivq, Kidder:1992fr, Damour2, Grishchuk}\,,
\begin{equation}
    \mathscr{E}(r,v) = \Psi(r) - \frac{1}{2}v^2
\end{equation}
with $\Psi(r)$ being the relative Newtonian gravitational potential. We can get a closed distribution function for such spherically symmetric dark matter distribution by following Eddington's inversion procedure\cite{Binney}, and for a power law spike, it comes out as 
\begin{equation}\label{3.3}
    f(\mathscr{E}) = \frac{\alpha(\alpha - 1)}{(2\pi)^\frac{3}{2}} \rho_{sp} \left( \frac{r_{sp}}{m_1}\right)^{\alpha}\frac{\Gamma(\alpha- 1)}{\Gamma(\alpha - \frac{1}{2})}\mathscr{E}^{\alpha-\frac{3}{2}}\,.
\end{equation}

%\( \Psi(r)\) is the relative Newtonian potential, which, close to the central IMBH, is approximated to \(\frac{m_1}{r}\). So, now eq~\ref{eq:3.03} looks like,

%\begin{equation}
%    \label{eq:3.04}
 %   \mathscr{E}(r,v) = \frac{m_1}{r} - \frac{1}{2}v^2
%\end{equation}

%Now, as the secondary is present in a dark matter halo environment due to the presence of the central binary, it will face a dynamical drag. And due to that fact, it will lose energy. We have to model the drag force to estimate this energy loss.
\par
 %After doing so, we can determine the energy flux radiated due to this halo-generated dynamical friction. Now the power radiated instantaneously has the form
%\begin{equation}
 %   \label{eq:3.05}
  %  \frac{dE}{dt} = - F(r,v) v
%\end{equation}

%But unlike the Elliptic case, where we could take an orbital average as it was a closed orbit, we would do something different. We would do something called a scattering average, where we would think a secondary comes from infinity up to an angular position \( \phi_{0}\), which is the position of the closest approach and then flies by, never coming to the same position.So our angular region of interest would be from \( \theta= 0\) to \( \theta = \theta_{0}\)
\par
%However, there are more complete stories when exploring physics related to DM medium. 
Whenever a stellar-mass object moves through the DM mini spike, it gravitationally interacts with the DM particles, and we can expect a ``drag" -a like process happening. Indeed such ``frictional" effects have a name for them, and they are called ``Dynamical friction" or ``gravitational drag." The gravitational drag force removes angular momentum from an object in orbital motion, causing it to spiral toward the orbit's center gradually. In a pioneering study, Chandrasekhar (1943) derived the classical formula of dynamical friction in a uniform collisionless background\cite{Chandra}, which has been applied to several astronomical systems. Examples include orbital decay of satellite galaxies orbiting their host galaxies \cite{Hashimoto:2002yt, Albada, Fujii:2005kw}, dynamical fates of globular clusters near the Galactic center\cite{Kim:2003hc}, galaxy formation within the framework of hierarchical clustering scenario \cite{Kauffmann:1993gv, Nagashima:2004gr}, and formation of Kuiper-belt binaries \cite{Goldreich:2002eq}. The drag formula based on perturbers moving straight in either a collisionless medium
or a gaseous medium depends on the Coulomb logarithm $\log\Lambda=\log(r_{max}/r_{min})$ where $r_{min}$ and $r_{max}$ are the cutoff radii introduced for avoiding a divergence of the force integrals. While many previous studies\cite{Binney, Kavanagh:2020cfn} conventionally adopted $r_{min}$ and $r_{max}$ as the characteristic sizes of the
perturber and the background medium, respectively, the choice of $r_{max}$ remains somewhat ambiguous for objects moving on near-circular orbits. However, in this paper we will focus on the hyperbolic tracks by the binaries. 
\par 
The frictional force has a form\cite{Chandra, Dokuchev, Ruderman, Rephaeli, Ostriker:1998fa}\,,
\begin{equation}
\label{3.4}
     F_{DF} = 4\pi m_2^2  \rho_{DM}(r) \xi(v) \frac{\log{\Lambda}}{v^2}\,.
\end{equation}
 There is a particular set of values for $\log \Lambda$. $m_2$ is the mass of the secondary. Here we take the one used in \cite{Kavanagh:2020cfn}, which is $\log \Lambda=\log \sqrt{\frac{m_{1}}{m_{2}}}$. In the subsequent sections, we have chosen our binaries to have masses $m_{1}=10^{3}M_{\odot}$ and $m_{2}=10 M_{\odot}$. This choice gives the value for the Coulomb logarithm to be around 1, which is the value we have chosen in our analysis\footnote{We should, however, emphasize that our results do not get affected much by this value of $\log\Lambda$, which is in line with the analysis in \cite{Cardoso:2020iji}.}. Moving on to the other term which needs an introduction is $\xi(v)$, which accounts for a fraction of dark matter particles having slower velocities than the orbiting masses. An estimate for the number of such particles can be sought by calculating the following integral.
\begin{equation}\label{3.5}
    \rho_{DM}(r) \xi(v) = 4\pi \int\limits_0^v v'^2 f\left( \Psi(r) - \frac{v'^2}{2}  \right) dv' 
\end{equation}
where the function $f$ defined in the integral above is expressed in (\ref{3.3}). There are a few salient points to state for the above expression. A thorough study in \cite{Kavanagh:2020cfn, Coogan:2021uqv} explored an analytical way to address the problem of modeling the evolution of the binary and the dark matter profile under some assumptions. These lead to a halo profile evolving in a fixed gravitational potential encoded in $\Psi(r)$\footnote{Interested readers are referred to\cite{Kavanagh:2020cfn, Coogan:2021uqv} for further details.}, which in turn simplifies the problem. The evolution of the dark matter density with time will be given by
\begin{equation} \label{3.6}
    \rho_{DM}(r,t)=4\pi\int^{v_{max}}_{0}v'^{2}f\Big(\Psi(r)-\frac{1}{2}v'^{2}\Big)dv'\,.
\end{equation}

%\textcolor{red}{While writing the above equation, a subtle point regarding $\xi(v)$ has been taken care of, which aligns with \cite{Kavanagh:2020cfn, Becker:2022wlo}. We focus on the part of $\xi(v)$ where the value saturates such that the velocity of the secondary object is always in the range $v=v_{orb}$, and we take the value approximately 1. Below this range $\xi(v)$ shows a variation with $v$ reported in \cite{Becker:2022wlo}. However, for analytic reasons, we choose our density to be static, i.e., $\rho(r,t)\sim \rho(r)$. {\bf Put it in thew section 5}}

 The functional form $f$ was defined as a spherically symmetric distribution function for the dark matter and depended on a relative Newtonian gravitational potential $\Psi(r)$ which can be approximated to $\sim \frac{m_{1}}{r}$ close to the central massive black hole \cite{Becker:2021ivq}. Putting all the expressions together (\ref{3.5}) has a closed-form expression:

\begin{equation}\label{3.7}
           F_{DF} =  K \int\limits_0^v  v'^2 \left( \frac{m_1}{r} - \frac{v'^2}{2} \right)^{\alpha - \frac{3}{2}}  dv' 
    \end{equation}
        with 
        \begin{equation}
          K = 4 \pi m_2^2 \frac{\log\Lambda}{v^2} \left\{  4\pi \frac{\alpha(\alpha-1)}{(2\pi)^\frac{3}{2}} \rho_{sp} \left( \frac{r_{sp}}{m_1}\right)^{\alpha}\frac{\Gamma(\alpha-1)}{\Gamma(\alpha-\frac{1}{2})}  \right\} \,. \label{K}
        \end{equation}
        
These expressions give us a perfect headstart to calculate the radiation fluxes and study the behavior of orbital elements of the binary in response to such media. We look into these in the subsequent sections.

%%%%%%%%%%%%%%%%%%%%%%%%%%%%%%%%%%%%%%%%%%%%%%%%%%%%%%%%%%%%%%%%%%%%%%%%%%%%%%%%%%%%%%%%%%%%%%%%%%%%%%%%%%%%%%%%%%%%%%%%%%%%%%%%%%%%%%%%%%%%%%%%%%%%%%%%%%%%%%%%%%%%%%%%%%%%%%%%%%%%%%%%%%%%%%%%%%%%%%%

\section{Calculation of Fluxes in the presence of dark matter} \label{sec3}
While calculating gravitational radiations from binaries in hyperbolic orbits, we assume while losing energy the orbit for the binaries remain Keplerian, and we do not have to deal with any additional corrections. This is much in line with the elliptic cases as well where we usually assume they lose energy on a timescale greater than the orbital timescale. Due to the dark matter medium, there will be a dissipative force $F_{DF}$, as given in (\ref{3.7}), and the total energy lost due to this force will be given by 
\begin{equation}
  P_{DM}(\phi_0)= \Big\langle \frac{dE}{dt}\Big\rangle=\int^{2\phi_{0}}_{0}\frac{d\phi}{\dot{\phi}}F_{DF}(r,v)v \label{4.1}
\end{equation}
One can use approximation methods to have an analytical handle to calculate this integral. To do that we expand the integrand of (\ref{3.7}) in powers of \( v'\)\,,

\begin{align}
    \begin{split}
        \label{eq:4.04}
          v'^2 \left( \frac{m_1}{r} - \frac{v'^2}{2}\right)^{\alpha - \frac{3}{2}} = v'^{2}\Big(\frac{m_{1}}{r}\Big)^{\alpha - \frac{3}{2}}\Big[1-\frac{rv'^{2}}{2m_{1}}\Big]^{\alpha-\frac{3}{2}}\simeq v'^{2}\Big(\frac{m_{1}}{r}\Big)^{\alpha - \frac{3}{2}}\Big[1-\Big(\alpha-\frac{3}{2}\Big)\frac{rv'^{2}}{2m_{1}}\Big]
         + \mathcal{O}(v'^6)\,.
    \end{split}
\end{align}

Then we get, 
\begin{equation}
    \label{eq:4.05}
    F_{DF} \approx  K \, \left [\, \frac{1}{3} \left( \frac{m_1}{r}\right)^{\alpha - \frac{3}{2}} v^3 - \frac{1}{10}\frac{\left( \left( \frac{m_1}{r}\right)^{\alpha - \frac{3}{2}} r (\alpha - 1.5)\right)}{m_1} v^5+\mathcal{O}(v^7)\right ]\,.
\end{equation}

If we rightly concern ourselves with binary speeds to non-relativistic only (which it should be), one can ignore all the higher orders in $v$ and restrict ourselves to only the leading order in the above expansion. Plugging in the expression for $v$, which we obtained in (\ref{eq:2.7}) we get a full \( \phi \) parameterized expression for the drag force as shown below, 

\begin{align}
    \begin{split}
        \label{4.06}
         F_{DF}\approx & \frac{16 \pi^2 m_2^2 v_o (b \sin(\phi_{0}))^{\frac{3}{2} - \alpha}}{3 \sin(\phi_{0})} \log\sqrt{\frac{1}{q}} \left\{  \frac{\alpha (\alpha - 1)}{(2\pi)^\frac{3}{2}}\rho_{sp} \left(\frac{r_{sp}}{m_1}\right)^\alpha \frac{\Gamma(\alpha -1)}{\Gamma(\alpha - \frac{1}{2})} m_1^{\alpha - \frac{3}{2}}\right\}  \\
        & \left[ \sqrt{(1+\cos^2(\phi_{0})-2\cos(\phi_{0})\cos(\phi - \phi_{0}))} (\cos(\phi - \phi_{0})- \cos(\phi_{0}))^{\alpha - \frac{3}{2}}\right]\,.
    \end{split}
\end{align}

This expression can be directly used in our $\frac{dE}{dt}$ integral:
\begin{align}
    \begin{split}
        \label{3.32}
        \frac{dE}{dt} & =  -v.F_{DF} \\
        & =  -\frac{16 \pi^2}{3} \left(\frac{v_o}{\sin(\phi_{0})}\right)^2  m_2^2 \log\sqrt{\frac{1}{q}} \left\{  \frac{\alpha (\alpha - 1)}{(2\pi)^\frac{3}{2}} \times \rho_6 \left(\frac{r_6}{m_1}\right)^\alpha \times \frac{\Gamma(\alpha -1)}{\Gamma(\alpha - \frac{1}{2})} \times m_1^{\alpha - \frac{3}{2}}\right\} \\
        & \hspace{0.5 cm} \Big\{ (b \sin(\phi_{0}))^{\frac{3}{2} - \alpha} \Big(1+\cos^2(\phi_{0})-2\cos(\phi_{0})\cos(\phi - \phi_{0}))\, (\cos(\phi - \phi_{0})- \cos(\phi_{0})\Big)^{\alpha - \frac{3}{2}}\Big\}
        \end{split}
\end{align}
 and upon performing the integral as mentioned in (\ref{4.1}), we get the total energy radiated as a function of $\phi_{0}$
 \begin{align}
    \begin{split}
        \label{eq:3.39}
          P_{tot}(\phi_o) & = P_{quad}+P_{DM} \\& =\kappa\, \frac{\sin^4(\frac{\phi}{2}) \sin^4(\frac{\phi}{2} - \phi_o)}{\tan^2(\phi_o) \sin^6(\phi_o)} \left [ 150 + 72 \cos(2\phi_o)+66 \cos(2\phi_o - 2\phi)-144(cos(\phi- 2\phi_o)+ \cos(\phi))\right] \\& - \lambda \int\limits_0^{2\phi_o} d\phi [(\cos(\phi - \phi_o) - \cos(\phi_o))^{\alpha - \frac{3}{2}}  (1 + \cos^2(\phi_o)  - 2 \cos(\phi_o) \cos(\phi - \phi_o))(\sin(\phi_o))^{-\alpha-\frac{1}{2}}]\,,
%        & - 2 \frac{G^2 M \rho k }{ v_o}  \cot(\frac{\phi_o}{2}) F[\frac{\phi_o}{2} , -\cos(\phi_o)\csc^4(\frac{\phi_o}{2})]\,,
    \end{split}
\end{align}
where $\kappa= -\frac{32  L^6}{45 c^5 b^8 \mu^4}\,.$
Here we have added the contribution coming purely from the GW radiation in the first line of (\ref{eq:3.39}) from \cite{Garcia-Bellido:2017knh, DeVittori:2012da}. The term proportional to $\lambda$ corresponds to the radiation coming out due to dynamical friction force generated by the dark matter medium. The $\lambda$ sitting in front of the integral is given in terms of the dark matter parameters:
\begin{align}
    \begin{split}
        \label{eq:3.34}
%        &     \left \langle P_{dm}(\phi_o)   \right \rangle = \\ 
%        &  - \lambda \int\limits_0^{2\phi_o} d\phi [(\cos(\phi - \phi_o) - \cos(\phi_o))^{\alpha_{spike} - \frac{3}{2}}  (1 + \cos^2(\phi_o) - 2 \cos(\phi_o) \cos(\phi - \phi_o))(\sin(\phi_o))^{-\alpha-\frac{1}{2}}] \\
%        & where \\ 
        & \lambda = v_o^2 b^{\frac{3}{2}-\alpha} \frac{16 \pi^2 m_2^2}{3} \log\sqrt{\frac{1}{q}} \left\{  \frac{\alpha (\alpha - 1)}{(2\pi)^\frac{3}{2}} \rho_{sp} \left(\frac{r_{sp}}{m_1}\right)^\alpha \frac{\Gamma(\alpha -1)}{\Gamma(\alpha - \frac{1}{2})} m_1^{\alpha - \frac{3}{2}}\right\}\,.
    \end{split}
\end{align}
The integral in (\ref{eq:3.39}) being performed over the allowed range of $\phi$. To solve the integral we have to resort to numerical means and in doing so we can look into the behavior of the flux as a function of $\phi_{0}$ and $\phi$ separately for the part coming purely from GW and the part coming from dynamical friction as shown in Fig.~(\ref{fig:Power_Fluxes}). Next we summarise our results below:

\begin{figure}
    \centering
    \includegraphics[scale = 0.35]{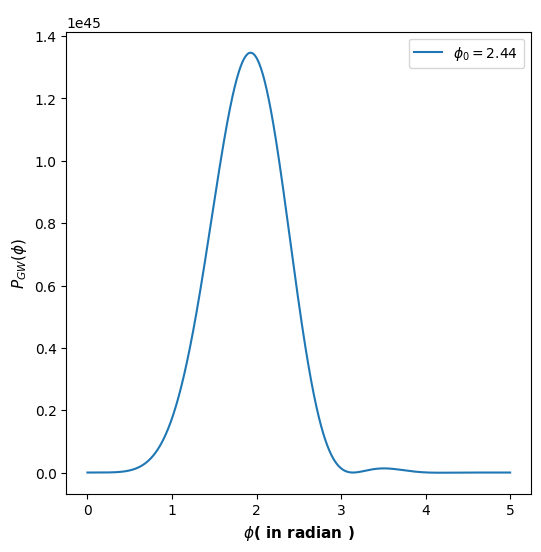}
    \includegraphics[scale = 0.35]{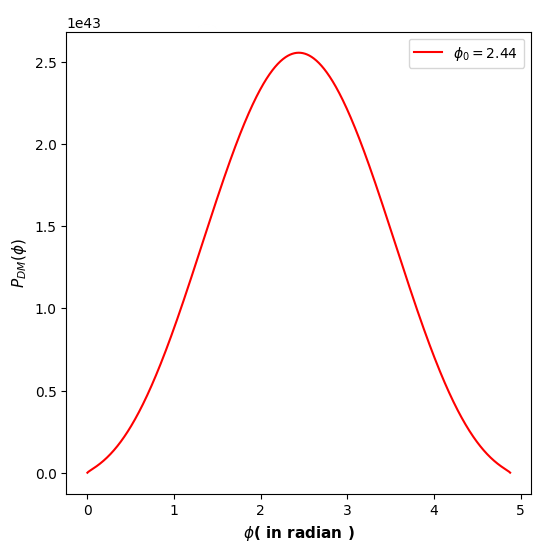}
    \caption{Total energy fluxes for the choice of $m_1= 10^{3}M_{\odot}, m_2= 10 M_{\odot}, \alpha=2.25, b=10^6 m, v_{0}=0.1c $ Left panel depicts purely GW contribution and right panel purely depicts the contribution due to dynamical friction. }
    \label{fig:Power_Fluxes}
\end{figure}

\begin{figure}[t!]
    \centering
    \includegraphics[scale = 0.4]{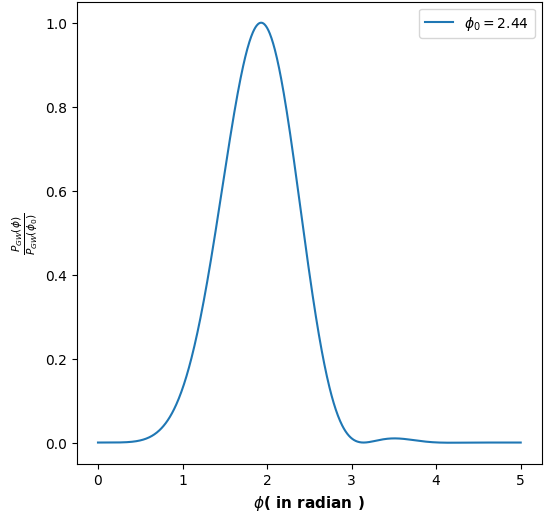}
    \includegraphics[scale= 0.4]{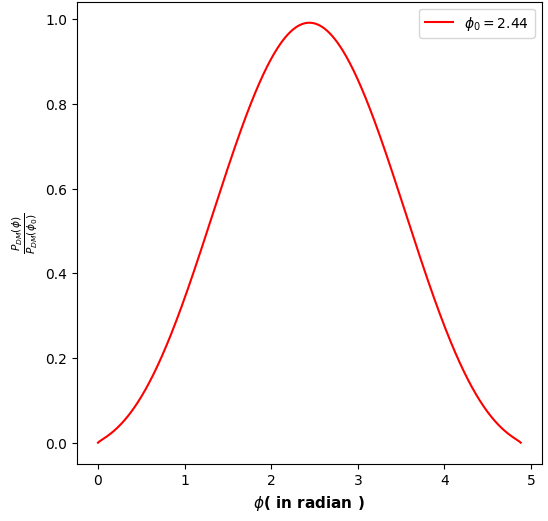}
    \caption{Normalized power fluxes for the choice of $m_1= 10^{3}M_{\odot}, m_2= 10 M_{\odot}, \alpha=2.25, b=10^6 m, v_{0}=0.1c $ Left panel depicts purely GW contribution and right panel purely depicts the contribution due to dynamical friction.}
    \label{fig:Relative_Power_Flux}
\end{figure}

To make a comparison, we focus on the ratio \( P(\phi)/P(\phi_0)\) for both Gravitational waves and the radiation in response to the Dynamical Friction of the medium. We chose $\phi_{0}=2.44$ for our analysis. The flux plots do give us a peak at $\phi=\phi_{0}$. When we put in the appropriate values of the parameters in the pre-factors, we see that the dark matter dynamical friction contribution is \( 10^{-2}\) (it is not obvious from plot) order lesser in magnitude. This is shown in Fig.~(\ref{fig:Relative_Power_Flux}). So, if we consider the net effect of radiation from such binaries in the presence of dark matter the dominant contribution comes from the gravitational radiation counterpart. 

\begin{figure}[b!]
    \centering
    \includegraphics[scale = 0.35]{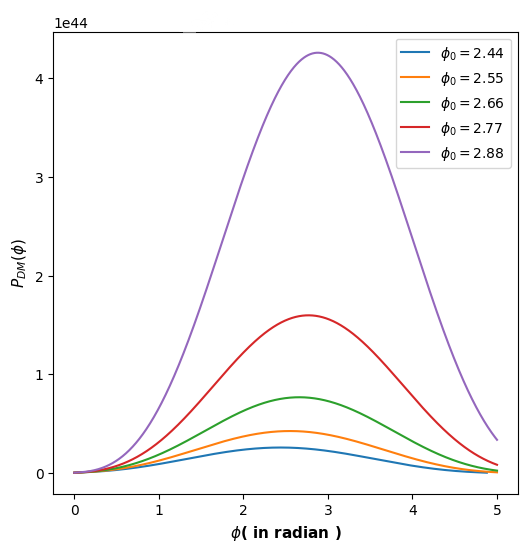}
    \caption{Fluxes due to dynamical friction for different choice of $\phi_0.$ Again we have set $m_1= 10^{3}M_{\odot}, m_2= 10 M_{\odot}, \alpha=2.25, b=10^6 m, v_{0}=0.1c.$}
    \label{fig:Comparative_DM_Fluxes}
\end{figure}

Finally, in Fig.~(\ref{fig:Comparative_DM_Fluxes}) we explore the nature of the flux due to the dynamical friction in the presence of the medium. Interestingly, we observe an increase in the peak of the flux with the increase in the angle of closest approach $\phi_{0}.$ This can be understood physically as with increasing $\phi_{0}$ the secondary is closer to the central black hole giving the secondary to interact with the dark matter halo for a longer period, hence the effect of the halo increases and so is the flux. 

 \section{Possible Observable signatures in GW signal} \label{barkingindex}
Having calculated the necessary fluxes associated with the GW emission in the presence of dark matter, including the effect of dynamical friction, we are in a good position to explore the possibilities of detecting such effects in the GW signal. In the context of pulsars, the so-called braking index $n_b$, has been a useful quantity to infer about the energy loss mechanism and these mechanisms are mainly mediated by GW emissions. Since pulsars can also spin-down through GW emission associated with asymmetric deformations, it is appropriate to take into account this mechanism in a model that aims to explain the
measured braking indices \cite{Ferrari, Ostriker}. The values of these braking indices can be $3.15\pm 0.03$, as reported in \cite{Archibald:2016hxz} for PSR J1640-46301, which was surprisingly the largest among the only eight of the $\sim$ 2400 known pulsars measured. This is expected from the pure magneto-dipole radiation model reported in \cite{Lyne '93, Lyne '96, Livingstone, Esp, Wel, Roy, Arch}. There have been many interpretations of this, such as those that propose that it may be due to the accretion of fallback material via a circumstellar disk \cite{Chen}, or via the quantum vacuum friction effect \cite{Coelho} or perhaps relativistic particle winds \cite{Xu, Wu} or modified canonical models \cite{Allen, EKsi}. Studies in \cite{Eps, Muslimov} have also indicated that it might be due to the effect of the change in the magnetic moment with time, either through a change in the surface field strength or the angle between the magnetic and spin axes. Taking a cue from this, \cite{De} showed that there could be possible combinations of gravitational and electromagnetic contributions to spindown, which could explain the measured braking indices.
\par
Once we have become familiar with the role of the braking index in inferring the characteristics of binaries, we can explore this in our context. Given the binaries (and the binaries are moving on a hyperbolic track), one can calculate the braking index using the following formula:
\begin{equation}
    n_{b}=\frac{\mathcal{F}\Ddot{\mathcal{F}}}{\dot{\mathcal{F}}^{2}}
\end{equation}
where $\mathcal{F}$ is the orbital frequency. In the context of elliptic orbits, $\mathcal{F}$ is related to the semi-major axis: $a$, via the relation $\mathcal{F}=\frac{1}{2\pi}\sqrt{\frac{M}{a^{3}}}\,.$ It is tricky to define the frequency for a hyperbolic orbit. The period is essentially infinite if a particle comes from infinity and goes to infinity after some scattering process following a hyperbolic track. However, there are many ways one can make a segue way out of this, like modeling the compact binaries after a semi-periodic source\cite{Brady:1998nj} or maybe focusing on a finite part of the hyperbolic track like going from point $x$ to $y$ and calculating the time taken for such a path. For a hyperbolic orbit, one can still redefine the usual orbital parameters like mean motion $n$ and the mean anomaly $\mathcal{M}$ as $\mathcal{M}=nt$ and $n=\sqrt{\frac{M}{-a^{3}}}$ which in turn will give us $\mathcal{F}=\frac{1}{2\pi}\sqrt{\frac{M}{-a^{3}}}$ and the period along the finite part of the track being $T=\frac{2\pi}{n}$
We also choose:
\begin{equation}
    r=\frac{a(e^{2}-1)}{1+e\cos\phi}
\end{equation}
with $e>1$ for the hyperbolic case. Furthermore, the energy of the orbiting particle in these tracks is expressed as $\frac{\mu M}{2|a|}$ ($>0$, as expected for an unbound orbit). Given these, we can express the braking index in terms of $a$ in the following way:
\begin{equation} \label{nb}
    n_{b}=\frac{5}{3}-\frac{2}{3}\frac{a\Ddot{a}}{\dot{a}^{2}}\,.
\end{equation}
The next step towards computing the braking index is to find $a, \dot{a}, \Ddot{a}$, and the method is straightforward. Assuming that while losing energy through radiation, we neglect the osculations of the orbit and the orbital timescale is such that we can use the Keplerian equations while taking the orbital average, we can calculate the dissipative forces acting on the binaries. We can model these dissipative forces by making them a function of $r$ and $v$ and then taking the average over the hyperbolic orbit. The energy and angular momentum corresponding to these dissipative forces can be calculated as follows\footnote{Later, we will convert the $t$ integral into a integral of $\phi$ using chain-rule, with the limits of integration changing from $t=(0,T)$ to $\phi=(0,2\phi_{0})$, where $\phi_{0}$ is shown in Fig.~\ref{gEOM} and the corresponding expression is given in (\ref{eq:2.4}). One can use the expression given in (\ref{eq:2.1}) to replace $\dot{\phi}$ in the integral.}: 
\begin{equation}
    \begin{aligned}
        & \Big\langle \frac{dE}{dt}\Big\rangle=-\int^{T}_{0}\frac{dt}{T}F(r,v)v\,,\\& \Big\langle \frac{dL}{dt}\Big\rangle=-\sqrt{M\,a(e^{2}-1)}\int^{T}_{0}\frac{dt}{T}\frac{F(r,v)}{v}\,.
    \end{aligned}
\end{equation}
Given an $F(r,v)$ we can compute $\Big\langle \frac{dE}{dt}\Big\rangle$ and $\Big\langle \frac{dL}{dt}\Big\rangle$. In our context, the dissipative effects that we will explore are the effects of dynamical friction with DM spike and the total orbital and angular momentum loss over the orbital timescale is given by
\begin{equation}
    \begin{aligned}
        & \frac{dE_{\text{orb.}}}{dt}=\Big\langle \frac{dE_{\text{GW}}}{dt}\Big\rangle+\Big\langle \frac{dE_\text{DM}}{dt}\Big\rangle\,, \\& \frac{dL_{\text{orb.}}}{dt}=\Big\langle \frac{dL_{\text{GW}}}{dt}\Big\rangle+\Big\langle \frac{dL_\text{DM}}{dt}\Big\rangle\,.
    \end{aligned}
\end{equation}
 To obtain the change in orbital parameters we take resort to the following equations, 
 \begin{equation}
 \begin{aligned}
    & E_{\text{orb.}}=\frac{\mu M}{2a}\Rightarrow \frac{da}{dt}=\frac{dE_{\text{orb.}}}{dt}/\frac{\partial E}{\partial a}\,,\\& \frac{de}{dt}=-\frac{e^{2}-1}{e}\Big(\frac{dE_{\text{orb.}}}{dt}/E_{\text{orb.}}+2\frac{dL_{\text{orb.}}}{dt}/L_{\text{orb.}}\Big)\,.
 \end{aligned}
 \end{equation}
 Given these equations and a particular model for the dissipative force (which is the dynamical friction force), we can calculate these evolutions. For our context, the dissipative force inspired is of the form:  $F_{0}r^{\gamma}v^{\delta}$ \cite{Becker:2022wlo}. Here, $F_0=4\pi m_2^2 r_{sp}^{\alpha}\log\Lambda.$ This follows from (\ref{3.4}). Upon using such a form for the dissipative force we can calculate $\dot{a}$ by
 \begin{equation}
 \begin{aligned}
  &   \dot{a}=\frac{dE_{\text{orb.}}}{dt}/\frac{\partial E}{\partial a}=-\frac{F_{0}}{\pi\mu}a^{k_{1}}(e^{2}-1)^{k_{1}-1/2}M^\frac{{\delta-2}}{2}\int^{2\phi_{0}}_{0}d\phi(1+e\cos\phi)^{-(2+\gamma)}(1+2e\cos\phi+e^{2})^{\frac{\delta+1}{2}}\,,\\& \dot{e}=-\frac{F_{0}}{2\pi}\frac{1}{e}\Big[(e^{2}-1)^{k_{1}+1/2}\frac{M^{\frac{\delta-2}{2}}}{\mu}a^{\tilde{k}_{1}}\int^{2\phi_{0}}_{0}d\phi(1+e\cos\phi)^{-(2+\gamma)}(1+2e\cos\phi+e^{2})^{\frac{\delta+1}{2}}\\& -(e^{2}-1)^{k^{\prime}_{1}+1/2}M^{\frac{\delta-3}{2}}a^{k^{\prime \prime}_{1}}\int^{2\phi_{0}}_{0}d\phi(1+e\cos\phi)^{-(2+\gamma)}(1+2e\cos\phi+e^{2})^{\frac{\delta-1}{2}}\Big]\,.
 \end{aligned}    
 \end{equation}
 To proceed further with these expressions, we expand these integrals around $e=1 $ to get an analytic grasp of these integrals (we leave the general case for future study). We do this by taking $e=1+\epsilon$ and then expanding in terms of $\epsilon$. We also consider a static profile, leaving the study of effect of evolution of the dar matter profile to braking index for future. Concentrating on the form of $F(r,v)$ for the dynamical friction force given in (\ref{3.4}), we can infer that $\gamma=-\alpha=-2.25$ (as defined below (\ref{model})\footnote{We consider only one value of $\alpha$ as at this point we are only interested in getting an order of magnitude estimate for the braking index. However, our analysis is valid for other values of $\alpha$  mentioned below (\ref{model}).} and $\delta=-2$ (this corresponds to setting $\xi(v)=1$ in (\ref{3.4}).). Using these values and also expanding in $\epsilon\,,$ we get
 \begin{equation}
 \begin{aligned} \label{eq:5.8}
 &    \dot{a}=-\frac{F_{0}}{2^{1/4}\pi \mu}\frac{a^{1/4}}{\epsilon^{1/4}\, M^{2}}\Big(1-\frac{\epsilon}{8}\Big)(6.23634-1.68719\sqrt{\epsilon})\,,\\& \dot{\epsilon}=-\frac{F_{0}}{2\pi}\frac{2^{3/4}\epsilon^{3/4}}{a^{3/4}}\Big[\frac{1}{\mu\, M^{2}}(6.23634-1.68719\sqrt{\epsilon})-\frac{1}{M^{5/2}}(0.51965-0.420448\sqrt{\epsilon})\Big]\,.
 \end{aligned}    
 \end{equation}
 As is evident from the definition of the braking index, $n_{b}$, given in (\ref{nb}), the above expressions are enough to infer the braking index for our scenario. 
 %\textcolor{blue}{Depends on the plot: Also, to get an order of magnitude estimate for the braking index when the dark matter density evolves, motivated by \cite{Becker:2022wlo}, we consider $\xi(v)\propto v^3$ \footnote{Here, we have assumed that the binary is moving with a non-relativistic velocity. Then, from (\ref{3.4}), it is evident that $F_{DF}$ is $\propto v\,.$ } to obtain an analytic handle. In this case, we will $\gamma=-\alpha=-2.25$ as before and $\delta=1\,.$ Then following the procedure mentioned above, we get}
%\textcolor{blue}{Put the correct equations
%\begin{equation}
% \begin{aligned}
% &    \dot{a}=-\frac{F_{0}}{\pi \mu}\frac{a^{1/4}}{m^{2}}\Big(\frac{1}{2^{1/4}\epsilon^{1/4}}-\frac{\epsilon^{3/4}}{8.2^{1/4}}\Big)(6.23634-1.68719\sqrt{\epsilon})\,,\\& \dot{\epsilon}=-\frac{F_{0}}{2\pi}\frac{2^{3/4}\epsilon^{3/4}}{a^{3/4}}\Big[\frac{1}{\mu m^{2}}(6.23634-1.68719\sqrt{\epsilon})-\frac{1}{m^{5/2}}(0.51965-0.420448\sqrt{\epsilon})\Big]\,.
% \end{aligned}    
% \end{equation}}\textcolor{red}{We have a problem with the value of the braking index for this case! that's why this is not included}
\begin{figure}[t!]
    \centering
    \includegraphics[scale = 0.35]{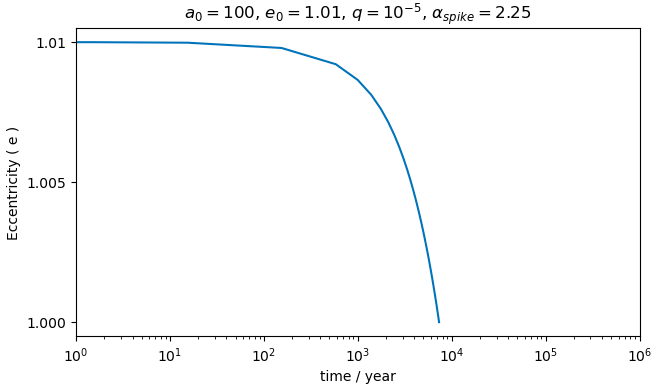}
    \includegraphics[scale = 0.45]{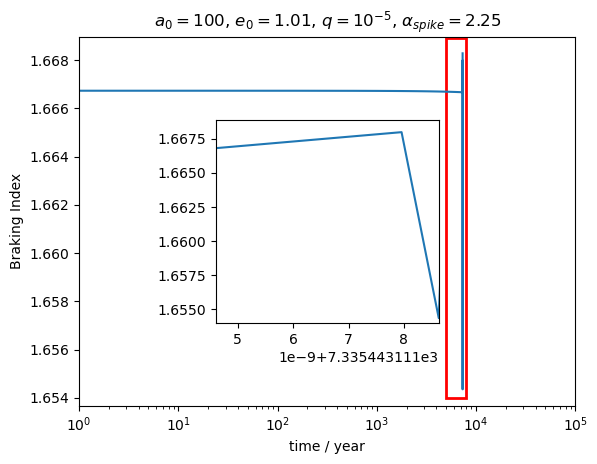}
    \caption{Plots showing the evolution of eccentricity and the evolution of the braking index for an initial eccentricity $e_{0}=1.01,$ initial semi-major axis $a_0=100$ and mass ratio $q=10^{-5}\,.$ Inside the figure in the right panel, we have included an inset showing zooming on the fall-off part of the braking index. The spiky behaviour is due to the limitation of our numerical precision. }
    \label{fig:new}
\end{figure}
\newpage
Now, we plot the braking index as a function of time. Fig.~(\ref{fig:new}) shows some interesting features worth mentioning. Early in the evolution, the braking index is constant with the saturation value being $\sim 1.666$, and this is the value of the index in ``vacuum", i.e., when the environmental effects or the spike effects are not there \footnote{The plot for the braking index that we show here shows a sudden spike when the value for the index starts to decrease on the timescale of $\sim 10^{4}$ yr. We emphasize that this ``spike" is due to some numerical artefact and not due to any other non-trivial physics.}. As they start to come closer together and enter the halo region, the braking index shows a fall, with an estimated time for the fall around $\sim 10^{4}$ yr. Firstly, the plot for eccentricity evolution serves as a consistency check that the $e$ remains in the vicinity of unity as assumed while deriving the relevant equations in (\ref{eq:5.8}). Secondly, it allows us to explore the effects of the environment at large times, which will lead to larger dephasing. The timescale for such effects to be predominant is roughly $\sim 10^{4}$ yr, nearly identical to the braking index plot. %\textcolor{blue}{say something about the $\xi=1$ and $\xi=v^3$ case.}\textcolor{red}{This is the $\xi=1$ case, the other is having some wrong numbers that's why they are not included}\par

The observability of the second derivative of frequency evolution $\mathcal{F}$ has been an interesting problem to deal with like the one reported in \cite{Robson:2018ifk} and so is the braking index for inspiraling binaries \cite{Renzo:2021aho}. These results have an indication that such effects might be possible to detect if the dissipative forces are strong enough. 
 \section{Calculation of Osculating elements}\label{sec4}

Now we focus on the study of the binary dynamics on a hyperbolic orbit in presence of dark matter halo. One quantity that we will be interested in studying is the change of eccentricity. We will do so in the framework of the perturbative Kepler problem. One of the interesting things in these perturbed Kepler problems is that Kepler’s third law still holds on a shorter timescale. In the absence of perturbations, we know from basic central force problems that the motion will have six orbital constants. However, due to the perturbing forces, these orbital elements, constant earlier, start to vary “slowly.” The method of osculating elements then gives us the equations governing these changes. For a more comprehensive treatment, the reader is requested to look into \cite{Poisson} for a textbook treatment.

 \begin{figure}[htb!]
    \centering
    \includegraphics[width=.55\linewidth]{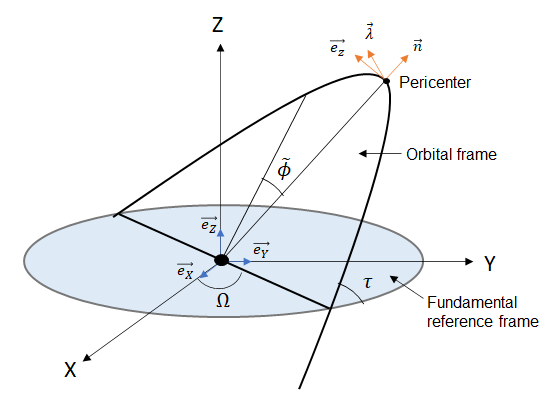}
     \caption{Orbital motion viewed in fundamental reference frame.}
    \label{fig:frame}
\end{figure}

\par

To tackle a Keplerian perturbation problem we use two coordinate systems- one is the orbital frame (OF) and the other one is the Fundamental frame (FF). Each frame comes with its own set of orthonormal basis vectors $\{e_{x},e_{y},e_{z}\}$ for the orbital frame while the fundamental frame has its own set $\{e_{X},e_{Y},e_{Z}\}$, each of them sharing a common origin and as usual the choice of the Fundamental frame is arbitrary. This is shown in Fig.~(\ref{fig:frame}). The two-body problem which we study in the presence of perturbation looks something like this:
\begin{equation}
    \vec{a}=\frac{m}{r^{2}}\hat{r}+\vec{f}
\end{equation}
with the perturbing force $\vec{f}$ \footnote{We like to remind the readers, that $\vec{f}$ is actually force per unit mass of the perturber. } being decomposed into a radial, cross-radial and another one perpendicular to the orbital plane in the form:
\begin{equation} \label{force}
    \vec{f}=\mathcal{R}\hat{n}+\mathcal{S}\hat{\lambda}+\mathcal{W}\hat{e_{z}}
\end{equation}
where, $\hat{n},\hat{\lambda},\hat{e}_{z}$ all depend on the various orbital parameters. These are all time dependent basis vectors in the OF. For the detail expressions of these basis vectors, interested readers are referred to \cite{Poisson}. Taking up true anomaly, $\tilde \phi=\phi-\phi_0 $, as our independent variable the equations governing the change of the orbital parameters are
\begin{align}
    \begin{split} \label{eq: 5.3}
        & \frac{dp}{d\tilde \phi}\simeq 2\frac{p^{3}}{M}\frac{1}{1+e\cos \tilde \phi}\mathcal{S}\,,\\& \frac{de}{d\tilde \phi}\simeq \frac{p^{2}}{M}\Big[\frac{\sin \tilde \phi}{(1+e\cos \tilde \phi)^{2}}\mathcal{R}+\frac{2\cos \tilde \phi+e(1+\cos^{2}\tilde \phi)}{(1+e\cos \tilde \phi)^{3}}\mathcal{S}\Big]\,,\\& \frac{d\iota}{d\tilde \phi}\simeq \frac{p^{2}}{M}\frac{\cos(\phi_0 +\tilde \phi)}{(1+e\cos \tilde \phi)^{3}}\mathcal{W}\,,\\& \sin\iota\frac{d\Omega }{d\tilde \phi}\simeq \frac{p^{2}}{M}\frac{\cos(\phi_0 +\tilde \phi)}{(1+e\cos \tilde \phi)^{3}}\mathcal{W}\,,\\& \frac{d\phi_0 }{d\tilde \phi}\simeq \frac{p^{2}}{eM}\Big[-\frac{\cos \tilde \phi}{(1+\cos \tilde \phi)^{2}}\mathcal{R}+\frac{2+e\cos \tilde \phi}{(1+\cos \tilde \phi)^{3}}\sin \tilde \phi\, \mathcal{S}-e\cot\iota \frac{\sin(\phi_0 +\tilde \phi)}{(1+e\cos \tilde \phi)^{3}}\mathcal{W}\Big]\,,\\& \frac{dt}{d\tilde \phi}\simeq \Big(\frac{p^{2}}{M}\Big)^{3/2}\frac{1}{(1+e\cos \tilde \phi)^{2}}\Big[1-\frac{1}{e}\frac{p^{2}}{M}\times \Big(\frac{\cos \tilde \phi}{(1+e\cos \tilde \phi)^{2}}\mathcal{R}-\frac{2+e\cos \tilde \phi}{(1+e\cos \tilde \phi)^{3}}\sin \tilde \phi\, \mathcal{S} \Big)\Big]\,.
    \end{split}
\end{align}
These equations as can be seen would have been constants had there been no perturbing force. They are controlled by the various components of the perturbing force and also parametrised by several orbital elements: the angle of pericenter (closest approach) $\phi_0 $; the angle of the ascending node $\Omega$; the eccentricity is $e$; the semimajor
axis is $a$; the semilatus rectum is $p$; and the inclination angle is $\iota$. 
We intend to extend this formalism for the case of hyperbolic orbits which is also a solution to the Kepler equations of motion. For this, we have to put in relations specific to the case of hyperbolic motions. The orbital elements that we are going to focus on are: impact parameter $b$, eccentricity $e$ of the track and the angle of closest approach $\phi_{0}$. Note that $\phi$ is the angle subtended between $m_{1}$ and $m_{2}$ with the horizontal line passing through $m_{2}$ and parallel to asymptotic line when $m_{1}$ is very far and $\phi_{0}$ is the angle when binaries are in closest approach or at pericenter. As mentioned in (\ref{eq:2.4}), the eccentricity is related to this angle $\phi_{0}$ through the relation $e=-\frac{1}{\cos\phi_{0}}.$  \par

Given this parametrization for hyperbolic orbit, the osculating equations implying the change in the eccentricity and $\phi_{0}$ due to the perturbations become the following \footnote{We want to point out that the other dynamical equations involving $\Omega$ and $\iota$ can also appear in principle, but since we are focusing on studying the change in eccentricity, we main focus on those two equations only and the other associated equations needed to solve it numerically.}, 
\begin{align}
    \begin{split} \label{fig:5.5}
        & \frac{dp}{d\tilde \phi}\simeq 2\frac{p^{3}}{M}\frac{1}{1+e\cos \tilde \phi}\mathcal{S}\,,\\& \frac{d\phi_0 }{d\tilde \phi}\simeq \frac{p^{2}}{e M}\Big[-\frac{\cos \tilde \phi}{(1+\cos \tilde \phi)^{2}}\mathcal{R}+\frac{2+e\cos \tilde \phi}{(1+\cos \tilde \phi)^{3}}\sin \tilde \phi \mathcal{S}-e\cot\iota \frac{\sin(\phi_0 +\tilde \phi)}{(1+e\cos \tilde \phi)^{3}}\mathcal{W}\Big]\,.\\
    \end{split}
\end{align}  
As, $e = - \frac{1}{\cos{\phi_{0}}}$ and $ b = \frac{ p }{e \sin{\phi_{0}}}$ for the hyperbolic orbits, evolution of the eccentricity $e$  and the impact parameter $b$ is govern by the following equations, 

\begin{align}
    \begin{split} \label{fig:5.6}
        &\frac{de}{d \tilde{\phi} } = - \frac{\sin{\phi_{0}} }{\cos^2{\phi_{0}} }\frac{d\phi_{0}}{d \tilde{\phi} }\,,\\& \frac{db}{d \tilde{\phi} } = -\cot{\phi_{0}}\frac{dp}{d \tilde{\phi}} - \frac{2b}{\sin{2\phi_{0}}} \frac{d\phi_{0}}{d \tilde{\phi} }\,.\\
    \end{split}
\end{align}

Apart from the conventional GW backreaction effects, the perturbations we will introduce are purely environmental effects like force due to the gravitational potential generated by the dark matter spikes and the dynamical friction and accretion associated with it. We discuss these sections step by step with relevant equations in the subsequent sections.

\subsection{Effect of gravitational potential due to the Dark Matter Minispike}
To proceed further, we consider a binary system where the mass of the central black hole is larger than that of its companion so that the reduced mass $\mu$ is approximately equal to the mass of the smaller companion. The barycenter position aligns itself with the position of the central black hole. By fixing a frame at the central black hole position or the barycenter position (they are both the same here), we can write out an equation of motion for the relative separation between the binaries in the following form:
\begin{equation} \label{eq: 5.7}
    \vec{a}_{G}=\frac{d^{2}\vec{r}}{dt^{2}}=-\frac{M_{eff}}{r^{2}}\hat{r}-\frac{F}{r^{\alpha-1}}\hat{r}\,.
\end{equation}
The parameters in the above equation depend on the parameters  of the dark matter minispike model as mentioned in (\ref{model}). They are defined as \cite{Eda:2014kra},  $$M_{eff}=M-\frac{4\pi \rho_{sp}r^{\alpha}_{sp}r^{3-\alpha}_{min}}{3-\alpha}\quad  \mathrm{and}\quad F=\frac{4\pi \rho_{sp} r^{\alpha}_{sp}}{3-\alpha}\,.$$ The unit vector $\hat{r}$ being directed from the central black hole to the smaller companion. 
\par
The first term in (\ref{eq: 5.7}) is the familiar gravitational interactions between two masses as mentioned and the second term purely originates due to the gravitational potential generated by the dark matter halo described a minispike model for our case. The term is like a perturbation to the usual Keplerian equation of motion \cite{Eda:2014kra}. Hence, we can use the formalism for osculating elements but ofcourse specialised to the case for hyperbolic orbits. As mentioned ealier, the orbital elements here are $\{e,b,\phi_{0}\}$ - all of them not being constant anymore due to the perturbation. 
\par
The above equation (\ref{eq: 5.7}) tells us the perturbing force spans the plane in which the binaries are moving and the components contributing to the perturbing force can be easily read off by comparing with (\ref{force}) and identifying $\hat{n}$ with $\hat{r}.$ Then from (\ref{fig:5.5}) and (\ref{fig:5.6}) we get,
\begin{align}
    \begin{split} \label{eq:5.7}
        & \frac{de}{d\Tilde{\phi}}=\frac{1}{\cos\phi_{0}}\frac{b^{3-\alpha}F}{M_{eff}}\frac{\sin^{4-\alpha}\phi_{0}\cos(\Tilde{\phi} )}{(\cos(\Tilde{\phi} )-\cos\phi_{0})^{3-\alpha}}\,,\\& \frac{db}{d\Tilde{\phi}}=\frac{b^{4-\alpha}F}{M_{eff}}\frac{\sin^{2-\alpha}\phi_{0}\cos(\Tilde{\phi})}{(\cos(\Tilde{\phi})-\cos\phi_{0})^{3-\alpha}}\,,\\& \frac{d\phi_{0}}{d\Tilde{\phi}}= - \frac{b^{3-\alpha}F}{M_{eff}}\cos\phi_{0}\frac{\sin^{3-\alpha}\phi_{0}\cos(\Tilde{\phi})}{(\cos(\Tilde{\phi})-\cos\phi_{0})^{3-\alpha}}\,,\\& \frac{dt}{d\Tilde{\phi}}=\sqrt{\frac{b^3 \sin^{3}\phi_{0}}{G M_{eff}|\cos\phi_{0}|^{3}}}\frac{\cos^{2}\phi_{0}}{(\cos(\Tilde{\phi})-\cos\phi_{0})^{2}}\Big[1-\frac{b^{3-\alpha}F}{M_{eff}}\cos\phi_{0}\frac{\sin^{3-\alpha}\phi_{0}\cos(\Tilde{\phi})}{(\cos(\Tilde{\phi})-\cos\phi_{0})^{3-\alpha}} \Big]\,.
    \end{split}
\end{align}

\subsection{Dynamical friction and accretion effects}
Given a binary moving through a medium (which for us is a dark matter medium) it will experience a drag or dynamical friction force. Having discussed the effects of dark matter as perturbing force, we can now focus on the effects of such matter medium on the overall dynamics in such hyperbolic tracks. To this end as we discussed in Section \ref{sec2} the ``frictional" force of the medium can be modeled after a force law which takes the form -

\begin{equation}
    \vec{f}_{DF} = - \frac{4\pi m_{2} \rho_{DM}\xi(v)\ln \Lambda }{v^{3}} \vec{v}
\end{equation}
where $f_{DF}$ is the perturbing force(per unit mass). At this point, we will make a simplifying assumption. We will choose the halo density to be static, i.e., $\rho(r,t)\sim \rho(r)$ by setting $\xi(v)=1$ following \cite{Dai:2021olt}. In general, as shown in \cite{Kavanagh:2020cfn, Becker:2022wlo}, the dark matter density will evolve in time, and the $\rho(r,t)$ can be found from (\ref{3.6}). But for our subsequent study, we will not consider this. We rather leave it for future investigation \footnote{For hyperbolic it will be interesting to figure out the nature of the phase space distribution function as mentioned in (\ref{3.6}) following a similar numerical analysis performed in \cite{Coogan:2021uqv, Kavanagh:2020cfn} in future.}. 

%But   \textcolor{red}{As discussed, the force law depends on $v$, which is the velocity of the body moving through the dark matter medium, the density of the medium: $\rho_{DM}$, and the speed of sound in the medium. While writing the above equation, a subtle point regarding $\xi(v)$ has been taken care of, which aligns with \cite{Kavanagh:2020cfn, Becker:2022wlo}. We focus on the part of $\xi(v)$ where the value saturates such that the velocity of the secondary object is always in the range $v=v_{orb}$, $v_{orb}$ is the orbital speed of the particle moving through the medium and we take the value approximately 1. One should note that we can consider a scenario where the binaries being initially in a hyperbolic track comes close and start to inspiral and then merge down\cite{Mouri:2002mc}. Hence, we safely choose that range of $\xi(v)$ where our binary orbits are fairly inspiralling down in circular orbits such that the value saturates at a constant value for a particular circular orbital velocity\cite{Becker:2022wlo}. Below this range, $\xi(v)$ shows a variation with $v$ reported in \cite{Becker:2022wlo} for the inspiralling black holes For analytic reasons, we further choose our halo density to be static, i.e., $\rho(r,t)\sim \rho(r)$.}
\par 
Not only can a body experience these frictional effects but there can also be non-trivial accretion effects. For that, considering the Bondi-Hoyle accretion effect and the velocity of the body in the medium to be in the range $v\gg c_{s}$ we can come up with a force law modeling the perturbating contribution\cite{Dai:2021olt}:
\begin{equation}
    \vec{f}_{acc} = - \frac{4\pi m_{2} \rho_{DM}B_{acc} }{v^{3}} \vec{v}
\end{equation}
where $B_{acc}$ is the term appearing from the model itself, we set it to unity for our analysis following \cite{Dai:2021olt}.

The dynamical friction and accretion force laws are in the direction of $\hat{v}$, which is just the unit vector of the binary velocities, namely, $\vec{v}=\dot{r}\hat{n}+r\dot{\tilde{\phi}}\hat{\lambda}$. Upon using these radial and cross radial directions, the radial and cross-radial components of the perturbing force can be again read off from the expressions by comparing with (\ref{force}), and they are: 
\begin{align}
    \begin{split} 
       & \mathcal{R} = 4 \pi  m_{2} r_{sp}^{\alpha} \rho_{sp} (\xi(v)\ln \Lambda  + B_{acc} ) \frac{b^{1-\alpha} }{M} \frac{ {\sin^{1-\alpha}}{\phi_{0}} \cos{\phi_{0}} \sin{\tilde{\phi}} ( \cos{\tilde{\phi}} - \cos{\phi_{0}}  )^{\alpha} }{ (1 + \cos^{2}{\phi_{0}} - 2\cos{\tilde{\phi}}\cos{\phi_{0} )^{3/2}}}\,,\\&
   \mathcal{S} = -4 \pi  m_{2} r_{sp}^{\alpha}\rho_{sp} (\xi(v)\ln \Lambda  + B_{acc} ) \frac{b^{1-\alpha} }{M} \frac{ {\sin^{1-\alpha}}{\phi_{0}} \cos{\phi_{0}} ( \cos{\tilde{\phi}} - \cos{\phi_{0}}  )^{\alpha + 1} }{ (1 + \cos^{2}{\phi_{0}} - 2\cos{\tilde{\phi}}\cos{\phi_{0} )^{3/2}}}\,.
    \end{split}
\end{align}

Once we have these components, we can plug them in the osculating equations (\ref{eq: 5.3}) and (\ref{fig:5.5}) and find the change in the orbital parameters:
\begin{align}
    \begin{split}\label{eq:5.10}
        & \frac{d\phi_0}{d\tilde{\phi}} = \frac{8\pi m_{2}\rho_{sp} r_{sp}^{\alpha} (\xi(v)\ln \Lambda  + B_{acc} )b^{3-\alpha}\sin^{3-\alpha}{\phi_0}\cos^2{\phi_0} (\cos{\tilde{\phi}} - \cos{\phi_0)}^{\alpha-1}\sin{\tilde{\phi}} }{M^{2}[  1 + \cos^2{\phi_0} - 2\cos{\tilde{\phi}}\cos{\phi_0} ]^{3/2}}\,,\\&\frac{de}{d\Tilde{\phi}} = - \frac{\sin\phi_{0}}{\cos^{2}\phi_{0}}\frac{d\phi_{0}}{d\tilde{\phi}}\,,\\& \frac{dp}{d\tilde{\phi}} = \frac{8\pi m_{2}\rho_{sp} r_{sp}^{\alpha} (\xi(v)\ln \Lambda  + B_{acc} )b^{4-\alpha}\sin^{4-\alpha}{\phi_0}\cos{\phi_0} (\cos{\tilde{\phi}} - \cos{\phi_0)}^{\alpha-2}}{M^{2}[  1 + \cos^2{\phi_0} - 2\cos{\tilde{\phi}}\cos{\phi_0} ]^{3/2} }\,, \\& \frac{db}{d \tilde{\phi} } = -\cot{\phi_{0}}\frac{dp}{d \tilde{\phi}} - \frac{2b}{\sin{2\phi_{0}}} \frac{d\phi_{0}}{d \tilde{\phi} }\,.
    \end{split} 
\end{align}

\subsection{GW Backreaction effects}
The system’s dynamics are dominated by the Newtonian gravitational attraction between the two bodies, and the radiation-reaction force creates a perturbation. The hyperbolic orbit solution emerges as a solution to the central force problem with $\frac{1}{r}$ potential. To include the effect of GW backreaction, we model it as a perturbing force to two body Kepler problems as \cite{Poisson}. Also, we are working in the post-Newtonian limit. To include leading order PN correction, we first have to modify the expression of the acceleration, which is given below \cite{Poisson}, 

\begin{equation}
\vec a_{GW} = \frac{8 m_1 m_2 }{5 c^5 r^3} \Big[ \Big(3v^2 + \frac{17\,M}{3r}\Big){r}^{.} \vec n - \Big(v^2 + \frac{3 M}{r} \Big) \vec v \Big]\,.
\end{equation}

The radial and cross-radial component of this force in terms of b, $\phi_{0}$ and $\tilde{\phi}$ in hyperbolic orbit case is given as- 

\begin{align}
    \begin{split}
        & \mathcal{R} = \frac{16 m^{3/2} m_{1} m_{2} }{15 c^{5} b^{9/2} } \frac{ ( -3 + 10 \cos{\tilde{\phi}} \cos{\phi_{0}} - 7\cos^{2}{\phi_{0}}) (\cos{\tilde{\phi}} - \cos{\phi_{0}} )^{3} \sin{\tilde{\phi}}  }{ \sqrt{- \sin^{9}{\phi_{0}} \cos^{3}{\phi_{0}}  } }\,,\\& \mathcal{S} = - \frac{64  m^{3/2} m_{1} m_{2} }{15 c^{5} b^{9/2} } \frac{ ( -1 + 5\cos{\tilde{\phi}} \cos{\phi_{0}} - 4\cos^{2}{\phi_{0}}) (\cos{\tilde{\phi}} - \cos{\phi_{0}} )^{4}  }{ \sqrt{- 64\sin^{9}{\phi_{0}} \cos^{3}{\phi_{0}}  } }\,.
    \end{split}
\end{align}

Once we have these components we can plug them in the osculating equations and find the change in the orbital parameters:
\begin{align}
    \begin{split} \label{eq:5.13}
        &\frac{d\phi_{0}}{d\tilde{\phi}} = - \frac{16  m^{1/2} m_{1} m_{2} }{15 c^{5} b^{5/2} } \frac{(\cos{\tilde{\phi}} - \cos{\phi_{0}} )\sin{\tilde{\phi}}}{ \sqrt{- \sin^{5}{\phi_{0}} \cos{\phi_{0}} } } \Big[ \cos{\tilde{\phi} }(  - 3 + 10 \cos{\tilde{\phi}} \cos{\phi_{0}} - 7 \cos^{2}{\phi_{0}} ) \\&\hspace{1.1cm}- (3/2)(  1 - 5 \cos{\tilde{\phi}} \cos{\phi_{0}} + 4 \cos^{2}{\phi_{0}} )(  2 \cos{\phi_{0}} - \cos{\tilde{\phi}} ) \Big]\,,\\&\frac{de}{d\Tilde{\phi}} = - \frac{\sin\phi_{0}}{\cos^{2}\phi_{0}}\frac{d\phi_{0}}{d\tilde{\phi}}\,, \\& \frac{dp}{d\tilde{\phi}} = - \frac{16 m^{1/2} m_{1} m_{2} }{5 c^{5} b^{3/2} } \frac{(\cos{\tilde{\phi}} - \cos{\phi_{0}} ) (  -1 + 5 \cos{\tilde{\phi}} \cos{\phi_{0}} - 4 \cos^{2}{\phi_{0}} ) }{ \sqrt{- \sin^{3}{\phi_{0}} \cos^{3}{\phi_{0}} } } \,,\\& \frac{db}{d \tilde{\phi} } = -\cot{\phi_{0}}\frac{dp}{d \tilde{\phi}} - \frac{2b}{\sin{2\phi_{0}}} \frac{d\phi_{0}}{d \tilde{\phi} }\,.\\
    \end{split}
\end{align}

\subsection{Results: Numerical Analysis and Inferences}
 Now that we have all the relevant equations for studying the binary dynamics, specifically the effects of these various forces due to the presence of dark matter on the rate of change of the eccentricity. Before discussing our result,  please note that we have used the fourth-order Runge-Kutta method to solve the coupled differential equations for each case separately, as mentioned in (\ref{eq:5.7}), (\ref{eq:5.10}) and (\ref{eq:5.13}). For each case, three coupled differential equations govern the evolution of eccentricity $e$, impact parameter $b$ and angle of closest approach $\phi_0$ w.r.t the true anomaly $\tilde{\phi}.$ Since we have taken true anomaly as our independent variable instead of time, all the plots are with w.r.t $\tilde{\phi}.$  The range of the true anomaly is from $\tilde{\phi} = -\phi_0$  to $\tilde{\phi} = \phi_0$ indicating the fact that the second object comes from infinity and means binaries after the encounter flies to infinity again. For the net effect, we first add the right-hand side of the (\ref{eq:5.7}), (\ref{eq:5.10}) and (\ref{eq:5.13}) and then sole the resulting coupled differential equation using the fourth-order Runge-Kutta method. %Also, while solving these equations, we have kept the closest approach between the central object and the secondary greater than the ISCO of the massive central BH.
\par
Now we discuss our findings. The results show some interesting features to comment upon, which we list below. Before that, we like to comment on the values of impact parameter $b$ that we have chosen for our numerical analysis. The values of $b$ are of $\mathcal{O}(pc)$, which is approximately $10^{16}m$, the reason being we are concentrating on the motion of binaries in the dark matter medium itself, which is distributed from $r_{min}$ (defined previously) to $r_{sp}$ which is approximately  of  $\mathcal{O}(10^{16}m)$.  
%\newpage

\begin{figure}
    \centering
    \includegraphics[width=.5\linewidth]{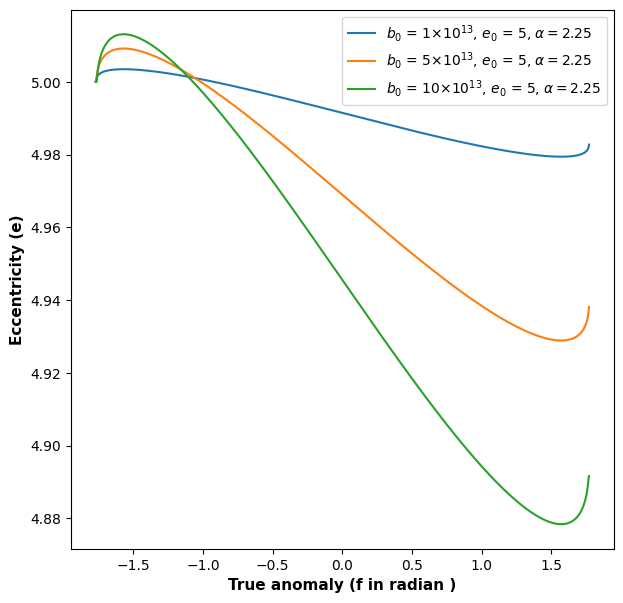}
    \caption{Effect of the gravitational potential of DM spike on the eccentricity for initial eccentricity $e_{0}$ = 5, $\alpha = 2.25$ and different values of initial impact parameter $b_{0}$ which is of the order $10^{13}\,m$.}
    \label{fig:test4}
\end{figure}

\begin{itemize}
    \item Let us first focus on the behaviour of eccentricity in the presence of gravitational potential due to the dark matter spike alone. This is shown in Fig.~(\ref{fig:test4}).  The figure shows a decrease in the eccentricity of the hyperbolic track. We started with an initial eccentricity $e_0=5$ and turned on the dark matter parameter $\alpha$ to look for the change in the eccentricity. The fall of eccentricity is evidently due to the dissipative nature of the dark matter medium itself, causing the binaries to lose energy in the form of radiation and eventually a change in eccentricity. Also from the Fig.~(\ref{fig:test5}) we can see if we increase the initial impact parameter for a fixed value of $\alpha,$ the change of $e$ is more pronounced. 

\begin{figure}[htb!]
    \centering
    \includegraphics[width=.5\linewidth]{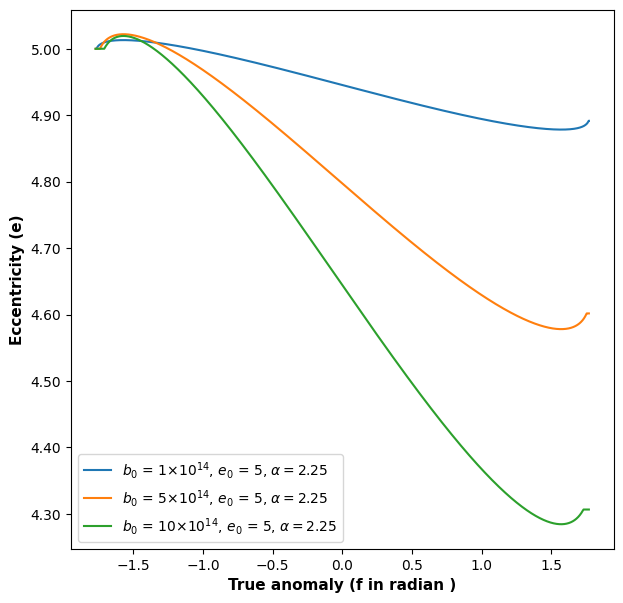}
    \caption{Effect of gravitational potential of DM spike for initial  eccentricity $e_{0}$ = 5, $\alpha = 2.25$ and different values of initial impact parameter $b_{0}$which is  of the order $10^{14}\,m. $}
    \label{fig:test5}
\end{figure}

Furthermore, as evident from the Fig.~(\ref{fig:test6}), the eccentricity falls off even more when we increase the dark matter parameter $\alpha$: $2.25, 2.33, 2.5,$ for a fixed value of impact parameter $b.$ The range of values for $\alpha$ that we have chosen, falls in the allowed range of constraints for $\alpha$ \cite{Boudaud:2021irr, Eda:2014kra}.

\begin{figure}[b!]
    \centering
    \includegraphics[width=.5\linewidth]{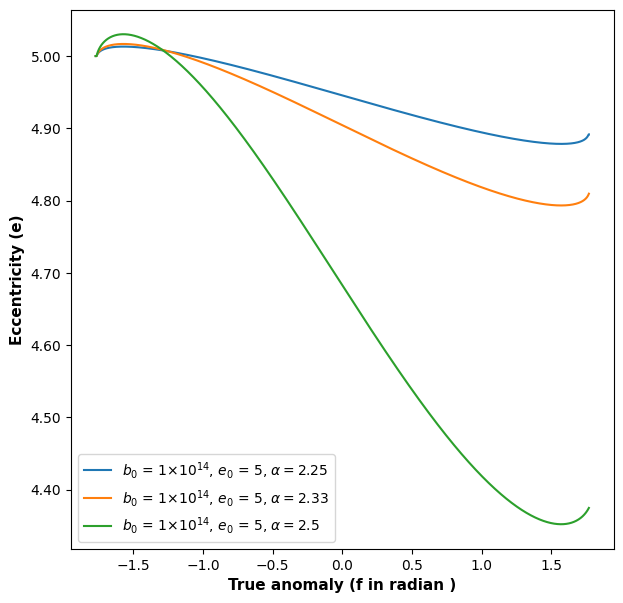}
    \caption{Effect of gravitational potential of DM spike for initial  eccentricity $e_{0} = 5$, initial impact parameter $b_{0} = 10^{14}\,m$ and $\alpha  = 2.25, 2.33\,\, \mathrm{and}\,\, 2.5. $}
    \label{fig:test6}
\end{figure}

    \item Note that the eccentricity is related to the angle $\phi_{0}$ through the relation $e=-\frac{1}{\cos\phi_{0}}$. Due to the perturbation arising from the dark matter profile, the eccentricity $e$ changes. Then this will, in principle, lead to a change in the angle $\phi_{0}$. Given a hyperbolic track, the angle $\phi_{0}$ is what the binaries make with the horizontal hyperbolic axis. A change in the angle $\phi_{0}$ would also cause a shift in the horizontal hyperbolic axis. We have numerically checked that the change in $\phi_0$ is negligibly small.

    \item Now, coming to the part where we study the effect of the drag force due to the presence of the dark matter medium, we see a stark change in results. Whereas previous studies have shown the dynamical friction to play an important role in the behavior of orbital parameters, we see a notable change here in this study. Eccentric binaries with elliptic orbits have shown an increase in eccentricity owing to the drag of the medium leading to the orbit getting wider \cite{Dai:2021olt}. However, we observe that the effect of the dynamical friction is subdominant compared to the effect due to the gravitational potential of the dark matter medium itself. Our analysis is done for several values of the initial impact parameter $b$, and we have observed negligible changes in the eccentricity arising due to the dynamical friction. As shown in the Fig.~(\ref{fig:7a}), for $b=10^{14}\,m,$ the change is of the $\mathcal{O}(10^{-4}).$

    \begin{figure}[htb!]
    \centering
 \includegraphics[width=.5\linewidth]{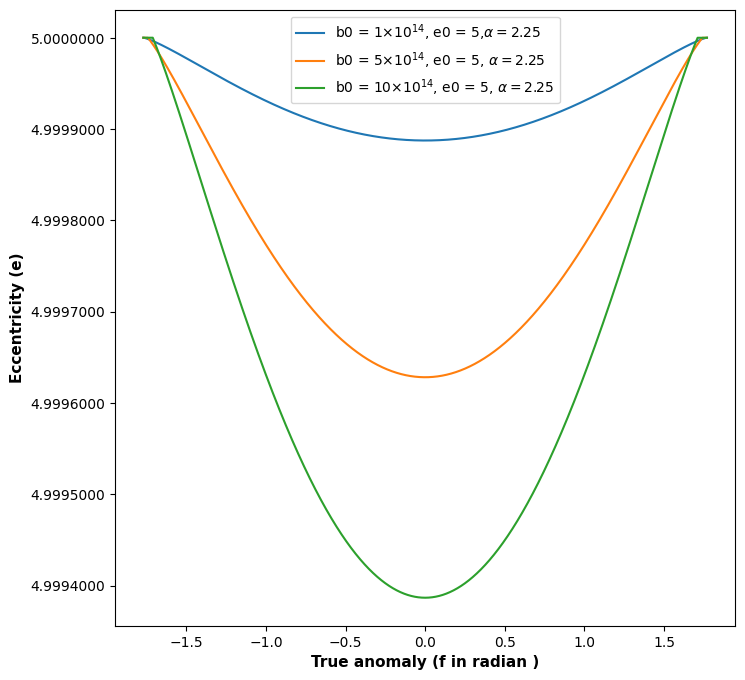}
    \caption{Effect for dynamical friction and accretion for initial eccentricity and impact factor $e_0$ = 5 , $b_0$ = $10^{14}\,m.$}
    \label{fig:7a}
\end{figure}

    \item Finally, we turn to the effect of the GW backreaction. For this case, we see a steep fall in the eccentricity values before it increases again, as seen from Fig.~(\ref{fig:test7}). The analysis is done for a fixed value of the initial impact parameter. A plausible reason for this dip in eccentricity can be accounted for by the increased backreaction effects when the binaries approach each other and scatter off, the backreaction effects being the strongest when they are closest. We can infer this as the variable controlling the parameters via the osculating equations is the true anomaly factor $\tilde{\phi}=\phi-\phi_{0}$. The sign of the true anomaly dictates whether the binary is closer to the secondary. This can be explained very logically since when the secondary is very far off $\tilde{\phi}$ is negative, and as it approaches the central massive black hole then $\tilde{\phi}$ approaches to zero and then start increasing again as the secondary is scattered off. 
    
    \item Also like the previous cases as $e=-\frac{1}{\cos\phi_{0}},$ change in $e$ implies in general a change in $\phi_{0}.$ But again this change in $\phi_0$ is negligible.

\begin{figure}[htb!]
    \centering
    \includegraphics[scale = 0.4 ]{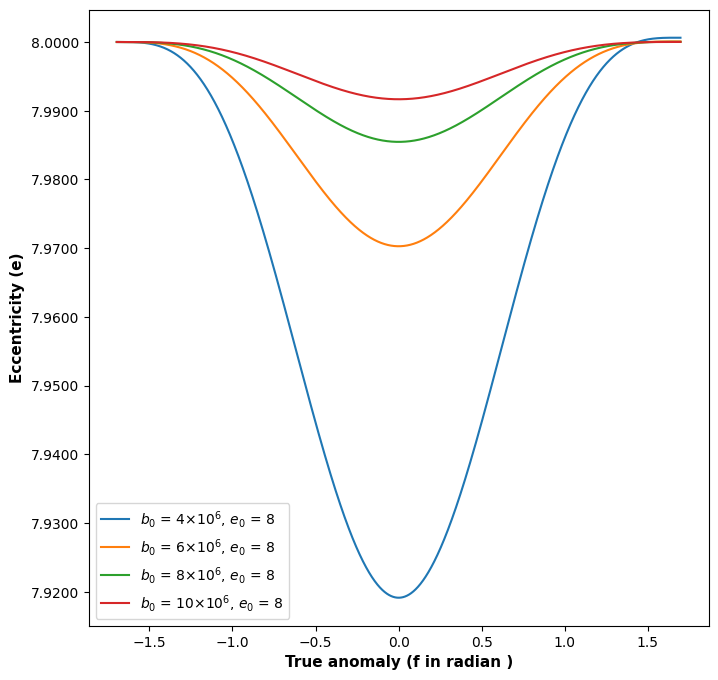}
    \includegraphics[scale = 0.4 ]{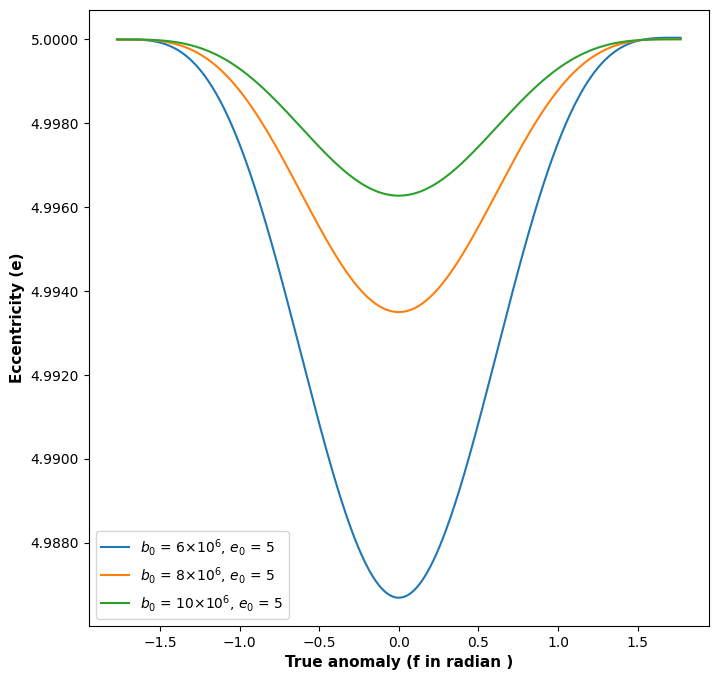}
    \caption{Effect for GW backreaction  for initial eccentricity and impact factor $e_0$ = 5 and 8 , $b_0$ = $10^6\,m.$}
    \label{fig:test7}
\end{figure}

 \item We can see by comparing Fig.~(\ref{fig:test7}) and Fig.~(\ref{fig:enter-labelnew}) that eccentricity changes due to GW backreaction is more when the initial impact parameter value is small. Also, if the initial eccentricity $e_0$ is large, then the rate of change of the eccentricity is also larger compared to the smaller initial eccentricity, as shown in Fig.~(\ref{fig:enter-labelnew}).

\begin{figure}[htb!]
    \centering
    \includegraphics[ scale = 0.4 ]{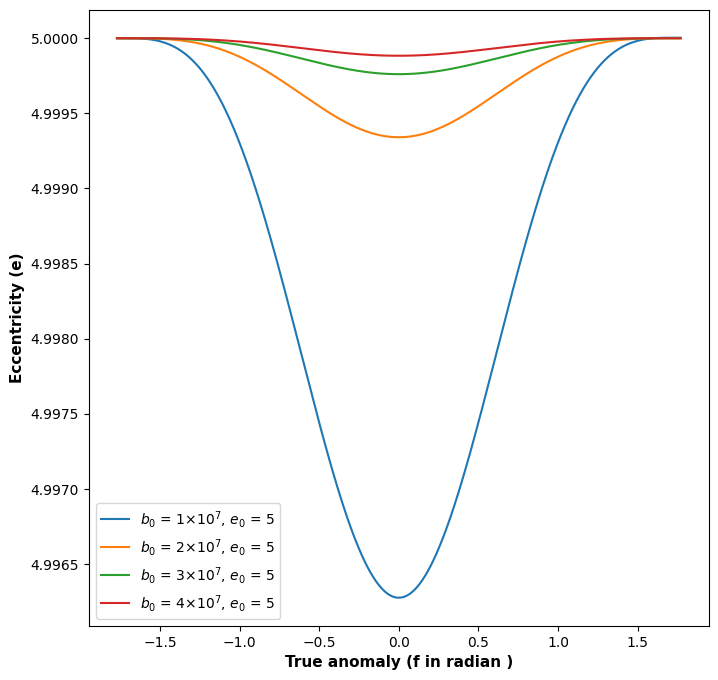}
    \includegraphics[ scale = 0.4 ]{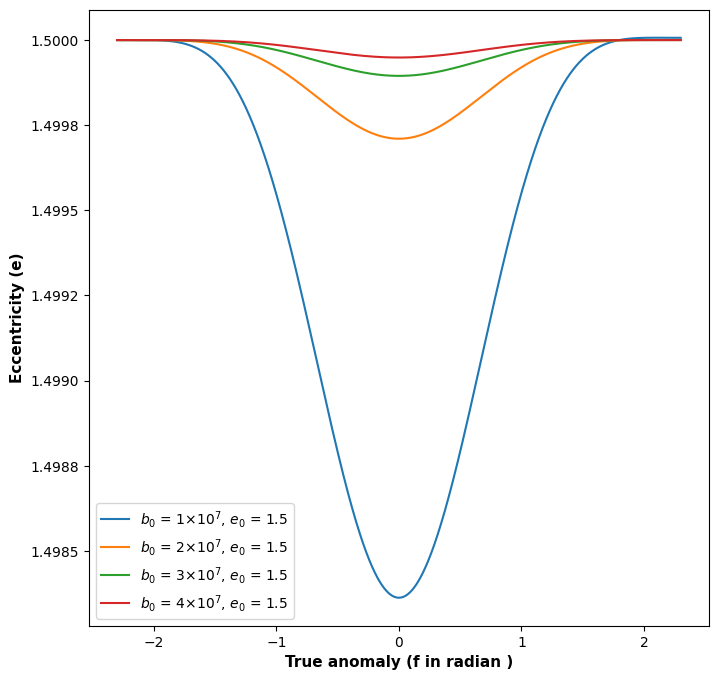}
    \caption{Effect for GW backreaction  for initial eccentricity and impact factor $e_0$ = 5 and 1.5 , $b_0$ = $10^7\,m.$}
    \label{fig:enter-labelnew}
\end{figure}

\item  \textit{Finally, we comment on the net effect when considering all the perturbing forces. The gravitational potential due to dark matter minispike and the GW backreaction force mainly contribute to the rate of eccentricity change. When the impact parameter $b$ is of $\mathcal{O}(10^{12}-10^{15})$ the initial decrease of $e$ is mainly governed by the gravitational force generated by the dark matter halo, but when $b$ is of  $\mathcal{O}(10^6-10^7)$then the dominant contribution comes from the GW backreaction. Also, the increase of $e$ at later times is entirely goverened by the GW backreaction effect.}

\begin{comment}
\begin{figure}[b!]
    \centering
    \includegraphics[width=.45\linewidth]{HEGWeVsfdiffb0106ande011.png}
    \caption{Effect for GW backreaction  for initial eccentricity and impact factor $e_0$ = 1.1 , $b_0$ = $10^6\,m.$}
    \label{fig:test9}
\end{figure}    
\end{comment}

\end{itemize}
%%%%%%%%%%%%%%%%%%%%%%%%%%%%%%%%%%%%%%%%%%%%%%%%%%
\subsection{The ``curious" case of baryons}
\textcolor{black}{In most cosmological models, numerical simulations regarding the formation of structures show an inner core of the dark matter halo, which survives and gives way to more intricate structures like ``subhalos" within their hosts. These subhalos boost gamma-ray production from dark matter annihilation and have also been known to enhance local cosmic ray production. However, all these estimates regarding such boosts and enhancements assume that baryon's gravitational effects on the dark matter substructure are negligible. However, studies in \cite{Cataldi, Chua} have been able to show that there are viable feedback effects that can significantly alter distributions. In \cite{Li:2022adj} show that given a dwarf galaxy setup, there could be dark matter heating due to baryons' adiabatic compression of dark matter. If $\rho$ is the dark matter density and $v_{d}$ is the velocity dispersion of the gas particles, we can define $F=\frac{\rho}{v_{d}^3}$ as the phase-space density. Unlike the $\rho$ for the dark matter density and $v_{d}$, this phase space density is unaffected in dark matter simulations. However, hydrodynamic simulations give a different picture for $F$, the trend being a gradual decrease with time due to stellar feedback. This decrease is on a factor of 10 as compared to dark matter simulations. The stellar feedback only in the galaxy's central region strongly affects the dark matter density. This is the region where the enclosed gas mass occasionally dominates that of the dark matter and is where the feedback most strongly affects the gas. At the end of the hydrodynamic simulations, the dark matter density at the smallest resolved radius becomes a factor of seven smaller than in the dark-matter-only simulations.}
\par 
\textcolor{black}{Given the particular shape of the halo profile, novel mechanisms like ``resonant heating" can introduce changes to the cusp profile. Dark matter simulations suggest such cusp profiles, consistent with predictions from the standard model. However, ``resonant heating" due to stellar feedback in hydrodynamic simulations turns these ``cuspy" profiles into a ``flat" core. One of the effects of this flattening is to reduce the efficiency of dynamical friction in the central regions. Again, this flattening is due to the stellar feedback (primarily arising out of baryons(stars and gas)). Such feedback may induce some heating in the globular cluster systems until stars stop forming. The velocity distribution is isotropic within the core and shows some radial anisotropy outside, which is not observed if massive gas clouds drive the mechanism. In that case, there are tangential anisotropies in addition to radial ones.}
\par 

\textcolor{black}{Given these effects, the response of dark matter distributions to baryons is topsy-turvy. While the above studies have shown a flattened curve to the profile, there are scenarios where the effects are otherwise. To resolve these discrepancies, theoretical arguments and simulations have proposed baryonic processes that can produce an expansion of the dark matter halo. Gas bulk motions, possibly supernova-induced in regions of high star formation activity, and the subsequent energy loss of gas clouds due to dynamical friction can transfer energy to the central dark matter component. In \cite{Maccio}, it has been pointed out that there is another (possibly more relevant) effect, namely the gas bulk motion can induce substantial gravitational potential fluctuations and a subsequent reduction in the central dark matter density. In \cite{Acevedo:2019gre}, it was shown that supernovae can eject matter from halos up to 100 km/s, but it has yet to be seen what impact this might have on dark matter profiles, nor how the addition of radiation pressure feedback might change things. So, while observations show evidence for flattened dark matter
density profiles up to L* galaxies, the question remains whether
there is enough energy input from baryons in more massive objects for these processes to be effective in altering the dark matter density profile of spiral galaxies with a dark matter mass of the order of $10^{11}-10^{12}$.}
\par 
\textcolor{black}{In light of these lessons, we focus on the density distribution and their related effects in our case, speaking of which we slightly change our profile following \cite{Balaji:2023hmy}. The new profile looks like the following:
\begin{equation}
    \rho_{DM}=\begin{cases}
    0 & \text{when } r < r_{min}\,,\\
        \rho_{sp}\Big(\frac{r_{sp}}{r}\Big)^{\alpha} & \text{when } r_{min}\leq r < r_{sp}\,,\\ \rho^{\text{cored}}_{\text{DM-HALO}}& \text{when } r\geq r_{sp}
        \end{cases} \label{modelb}
\end{equation}
where the cored halo profile has the following structure:
\begin{equation}
    \rho^{\text{cored}}_{\text{DM-HALO}}=\begin{cases}
        \rho^{\text{NFW}}_{\text{halo}}(r_{c})\Big(\frac{r}{r_{c}}\Big)^{-\gamma_{c}} & \text{when } r < r_{c}\,,\\ \rho^{\text{NFW}}_{\text{halo}}& \text{when } r\geq r_{c}
        \end{cases} \label{modela}
\end{equation}
}
\textcolor{black}{where $\rho^{\text{NFW}}_{\text{halo}}=\rho_{s}(\frac{r}{r_{s}})^{-1}(1+\frac{r}{r_{s}})^{-2}$ is the ubiquitous NFW profile with $r_{s}=18.6$ kpc and $\rho_{s}=\rho_{\odot}(R_{\odot}/r_{s})(1+R_{\odot}/r_{s})^{2}$, where $R_{\odot}=8.2$ kpc is the Sun position and $\rho_{\odot}=0.383$ GeV/$\text{cm}^{3}$, being the local DM density. Upon ``correcting" the profile taking into the baryonic effects as discussed before into consideration, we observe the following features:}
\begin{figure}[htb!]
   
     \includegraphics[ scale = 0.3 ]{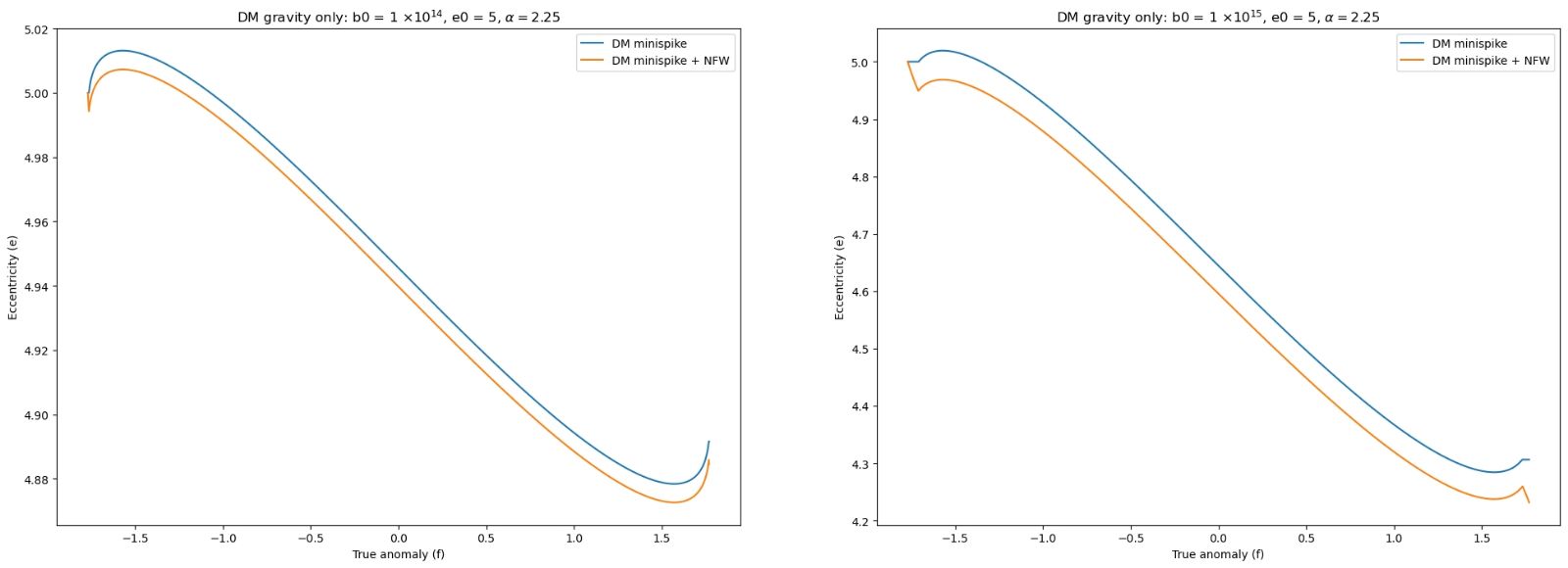}
    \caption{The above figures show the variation of eccentricity as a function of true anomaly parameter for different values of impact parameter $b_{0}$. The blue one is the one with the presence of spike only and the orange one is the one with an additional halo region covering it. The set of values we take are: $M_{\text{DM}}=10^{6}M_{\odot},M_{\text{BH}}=10^{3}M_{\odot}, r_{\text{s}}=23.1 \text{pc}, 
 \rho_{\text{s}}=3.9 \times 10^{-19}\text{kg}/\text{m}^{3}, r_{\text{sp}}=0.54 \text{pc}, \rho_{\text{sp}}=226 M_{\odot}/\text{pc}^{3}$ and $r_{\text{min}}=\frac{6GM_{\text{BH}}}{c^2}\,.$}
    \label{fig:enter-label}
\end{figure}

\begin{itemize}
 \item  \textcolor{black}{The eccentricity of the orbit shows a sharper fall-off than the one without the halo, as can be seen in Fig.~(\ref{fig:enter-label}). This can be understood logically since the spike and halo now dampen the hyperbolic binaries, leading to a greater fall for the eccentricities.}

\begin{figure}[htb!]   
    \includegraphics[ scale = 0.35 ]{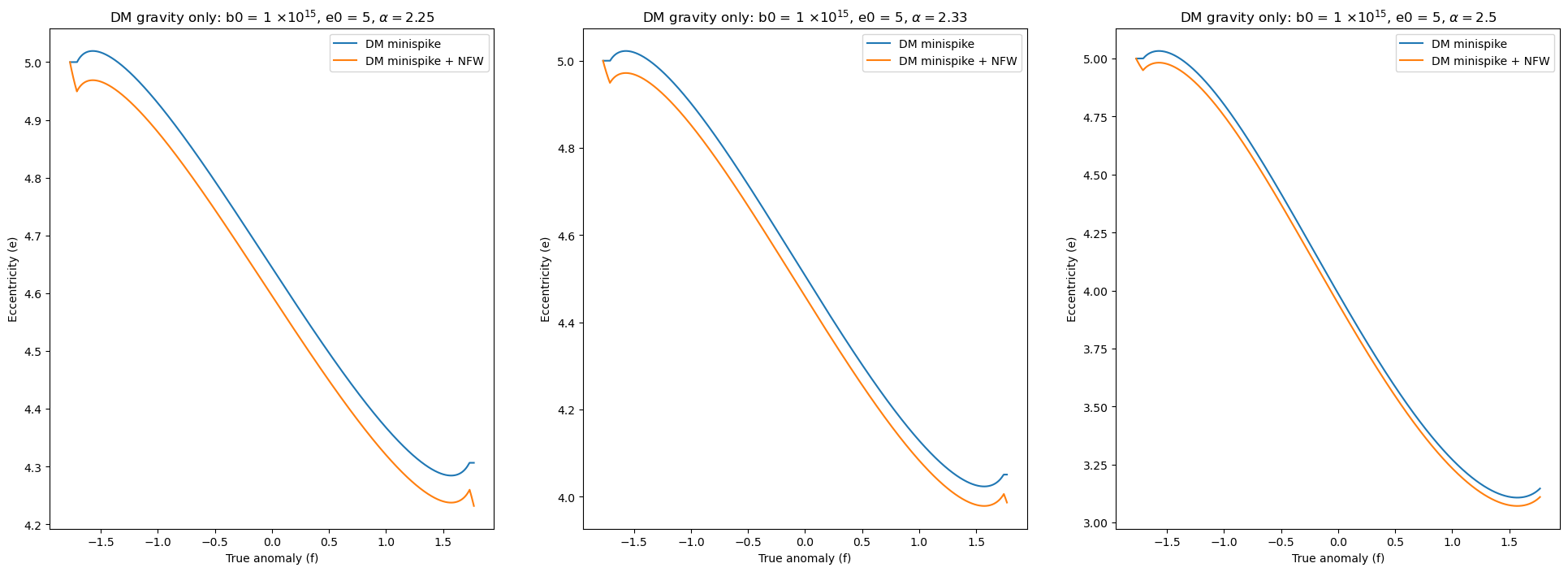}
    \caption{Plots showing the behavior of eccentricity as a function of impact parameter for different values of $\alpha$, the dark matter parameter index. The blue one is only for the spike profile and the orange one is the one with an additional halo region covering it. The set of values we take are: $M_{\text{DM}}=10^{6}M_{\odot},M_{\text{BH}}=10^{3}M_{\odot}, r_{\text{s}}=23.1 \text{pc}, 
 \rho_{\text{s}}=3.9 \times 10^{-19}\text{kg}/\text{m}^{3}, r_{\text{sp}}=0.54 \text{pc}, \rho_{\text{sp}}=226 M_{\odot}/\text{pc}^{3}$ and $r_{\text{min}}=\frac{6GM_{\text{BH}}}{c^2}\,.$}
    \label{fig:enter-label1}
\end{figure}
 
  \item  \textcolor{black}{Another interesting feature we observe in our profile is when we vary the dark matter parameter $\alpha$. As we increase the index - the variation being in the permissible range of 2.25 to 2.5, we see the two profiles merging. This is evident from Fig.~(\ref{fig:enter-label1}). To remind the readers again, these two profiles are the dark matter spike with the halo and the one without the halo. }
\end{itemize}
\newpage
\textcolor{black}{\subsubsection{Including the ``annihilation" region}}
\textcolor{black}{We can also do better and include some other effects too. As can be rightly inferred, the spike radius saturates due to DM annihilations. One can look into the physics of such annihilation in our orbital parameter behaviour. These annihilations happen in a region which is the innermost of this halo plus spike domain, weakening the density profile there. When considering models like the WIMP ones, the annihilation cross section $\langle \sigma v \rangle$ is constant, flatting the dark matter profile. Let $r=r_{\text{anni}}$ be this ``plateau" region and $\rho_{\text{anni}}$ being the corresponding density, then \cite{Shapiro:2016ypb}
\begin{equation}
    \rho_{\text{anni}}=\frac{m_{\xi}}{\langle \sigma v \rangle T} 
\end{equation}
where $m_{\xi}$ is the dark matter particle mass and $T$ is the galaxy age. }
\par
\begin{figure}[t!]
    
    \includegraphics[ scale = 0.35 ]{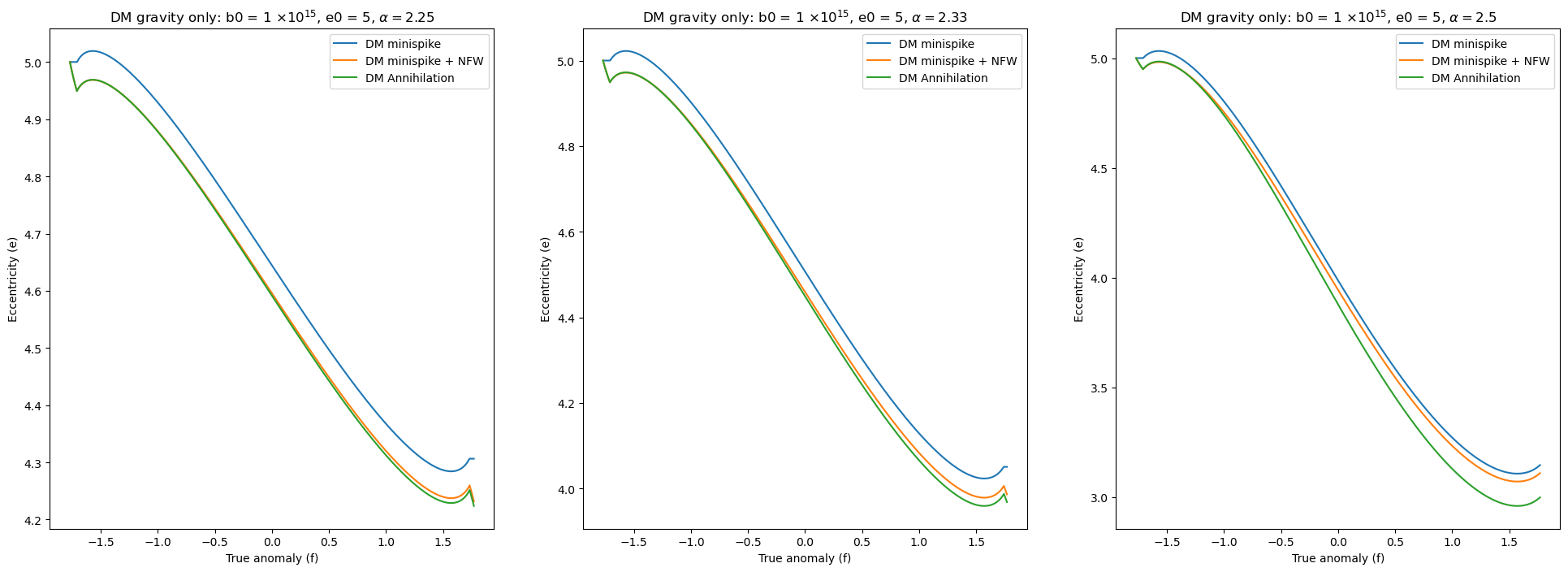}
    \caption{Variation of eccentricity upon the addition of the ``annihilation" region (green curve). We take $r_{anni}=3.1\times 10^{-3}\text{pc}$ and $\rho_{\text{anni}}=1.7\times 10^{8}M_{\odot}\text{pc}^{-3}$.  The set of values we take are: $M_{\text{DM}}=10^{6}M_{\odot},M_{\text{BH}}=10^{3}M_{\odot}, r_{\text{s}}=23.1 \text{pc}, 
 \rho_{\text{s}}=3.9 \times 10^{-19}\text{kg}/\text{m}^{3}, r_{\text{sp}}=0.54 \text{pc}, \rho_{\text{sp}}=226 M_{\odot}/\text{pc}^{3}$ and $r_{\text{min}}=\frac{6GM_{\text{BH}}}{c^2}\,.$}
    \label{fig:enter-label2}
\end{figure}

 
\textcolor{black}{A possible explanation for these plateaus was given in \cite{Vasiliev:2007vh}, due to dark matter (DM) particles moving in strictly circular orbits around the central black hole. In \cite{Vasiliev:2007vh}, it was shown that if the dark matter distribution is isotropic (which was likely so in all the simulations and analysis), the density shows a steep rise and eventually forms a cusp and not a plateau. The density varies as $\sim r^{-1/2}$ and the cusp is maintained as the particles outside
$r=r_{\text{anni}}$ continue to contribute to the density inside $r=r_{\text{anni}}$. The distinction between a plateau and a cusp has important observational consequences \footnote{\textcolor{black}{For a brief review see \cite{Shapiro}.}} and hence an analysis including such subtle but non-trivial physics is somewhat lucrative.
\begin{equation}
    \rho_{DM}=\begin{cases}
    0 & \text{when } r < 2r_{\text{S}}\,,\\
    \rho_{\text{anni}}\Big(\frac{r}{r_{\text{anni}}}\Big)^{-0.5} & \text{when } 2r_{\text{S}}\leq r < r_{\text{anni}}\,,\\
        \rho_{sp}\Big(\frac{r_{sp}}{r}\Big)^{\alpha} & \text{when } r_{\text{anni}}\leq r < r_{sp}\,,\\ \rho^{\text{cored}}_{\text{DM-HALO}}& \text{when } r\geq r_{sp}\,.
        \end{cases} \label{model1}
\end{equation}
}
\textcolor{black}{Upon including these effects, the behaviour in eccentricity shows the usual trend as demonstrated previously, meaning that it shows a sharper fall-off with the changes of $\mathcal{O}(10^{-2})$ than the one without them. Also, as seen in the case of halo, the profiles show a typical merging behaviour at the high values of $\alpha$. This is shown in Fig.~(\ref{fig:enter-label2}).}

\section{Conclusion}\label{sec9}\label{sec6}
In this work, motivated by its astrophysical significance and the prospect of its detection by future GW detectors and PTA \cite{Dandapat:2023zzn}, we studied the dynamics of hyperbolic encounters in the presence of the dark matter medium. Our approach to studying this involved computing radiation flux and the orbit dynamics of these encounters by treating the effect of the medium as a perturbative term in the Keplerian equations of motion. \textit{We observe that the power radiation due to the GW dominates over the radiation due to the dynamical friction generated by the dark matter halo. Next, while computing the change of eccentricity by computing various osculating elements, we observe again the effect of dynamical friction force felt by the secondary while interacting with the dark matter halo, and the effect of accretion force is negligible. But, the effect of the gravitational potential generated by the dark matter halo and GW backreaction effect is significantly more.} There's a genuine fall-off for eccentricity for the binaries due to these two effects for a certain time, while there is an eventual increase of the eccentricity due to the GW backreaction effect at late times. \textit{Interestingly, we observe that the change in eccentricity is more due to the dark matter potential for larger values of impact parameters. On the other hand, change in the eccentricity due to the GW backreaction increases with the decrease of impact parameter for a given value initial eccentricity.} This can be attributed to the fact that, as we increase the initial impact parameter the secondary object gets to interact with the dark matter halo more, thereby increasing the effect of the dark matter potential in the change of the eccentricity. On the other hand, as we decrease the impact parameter, the GW backreaction effect is larger.
\textit{These results are encouraging as the effect of the dark matter potential, which also depends on the parameter of the underlying model, provides a change in eccentricity and consequently can be constrained by observing such flyby events in future detectors.} \textcolor{black}{Furthermore, we extended our analysis by considering the effects of baryonic matter that may be present inside the halo. As discussed, it leads to the change of the profile, and we have considered one such model and analyzed the evolution of eccentricity. Curiously, it shows a sharper fall-off due to the extra damping provided by the presence of such matter.} 
\par
However, this analysis paves the way for only so much that can be done in these contexts. In doing the above study, we have made a number of simplifications, the foremost being the PN order in which it is to be computed. We have neglected the higher-order Post-Newtonian effects for these systems. For the flux calculation, we have only used the leading order PN results. The shape of the halo, which is taken here to be spherically symmetric, also needs to be relaxed, but we leave that for future work. Furthermore, while computing the osculating elements, we have neglected the evolution of the dark matter density profile. We have used only the static profile. But in general, the dark matter density profile evolves over time which is captured through the phase space distribution function \cite{Coogan:2021uqv, Kavanagh:2020cfn}. \textcolor{black}{Also, the dark matter profile mentioned may not remain same during the entire duration of binary motion and merger \cite{Phinney:2001di}. It will be interesting to find a way to setup our computation done in this paper and investigate the evolution of eccentricity for that case.} We like to include this in our future analysis as this will help us to make contact with a more realistic scenario. Furthermore, generalizing the analysis to include spinning binaries in this halo is also an interesting avenue to pursue in the future. Also, throughout our study, we have only focused on NFW  dark matter profile. Recently, there have been other interesting dark matter profiles \cite{Cardoso:2022whc}. It will be interesting to extend our study to those cases. Last but not least, it will also be interesting to use the tools from the effective field theory to study this problem, perhaps along the line of \cite{Huang:2018pbu, Bhattacharyya:2023kbh}. \par 
We have also looked for observational tools like braking index. The braking index is a useful tool for binaries at large separations and also in regimes where the environmental effects are dominant. We find that, it starts from a constant value when the bianries are in vacuum and start to decrease when the come close together inside the dark matter halo region. %\textcolor{blue}{We have also compared the two cases when the dark matter profile is static with the scenario when it has a particular time dependence coming through the $\xi(v)$ as defined in the main text.}
The study of eccentricity evolution complements it while also helping us to infer about environmental effects. Of course, our dissipative model was a simplistic one, $F=F_{0}r^{\gamma}v^{\delta}$, with more general forces having a dependence on spin, tidal deformability, etc. If we want to include more complex effects like halo feedbacks \cite{Kavanagh:2020cfn, Coogan:2021uqv}, there can be a possibility of a new equilibrium\cite{Coogan:2021uqv} from the phase parameterization they develop. Eventually, one should go for studying relativistic and post-Newtonian effects in this framework, and this is something we would definitely like to take over in some future study.
\par
Our analysis of osculating elements provides us with a platform to construct the GW waveform \cite{Dandapat:2023zzn}. By solving the osculating equation, we can solve for the evolution of the binary's orbital parameters, which in turn can be used to construct a waveform \cite{Dai:2021olt}. Then it will give us a route to understand and probe into the effects of dark matter contributions in such systems and possibly puts some constraints on the parameters using the GW data. In fact, following the analysis of \cite{Coogan:2021uqv}, one can possibly comment on the dark matter environment by analyzing the dephasing of such GW waveform. We are currently in the process of constructing such a waveform for flyby events in the presence of dark matter and hope to report on that issue in the near future. \textcolor{black}{This will also help us to put constraints on the dark matter using multi-messenger astronomy similar to what has been done for the circular orbit \cite{Ghoshal:2023fhh}. Note that the analysis presented in this paper is valid for any mass ratios of the binary. It will be interesting to analyze the dephasing of gravitational waveform in the extreme mass-ratio limit \cite{Rahman:2023sof}. This will be an interesting study in the context of gravitational wave phenomenology using future detectors like LISA. } 
\par
To summarize, in this work, we provide a setup for the realistic modelling of binaries in hyperbolic orbits surrounded by dark matter, which can be useful to probe into some new information in future detectors capable of detecting flyby events. Detecting dark dresses would have an impact beyond astrophysics and cosmology since their density profiles depend on the dark matter's fundamental properties. Hence their detection (both from the closed and the open orbits) would therefore provide a powerful probe of the particle nature of dark matter.
%\newpage

\section*{Acknowledgements}

Research of A.C. is supported by the Prime Minister's Research Fellowship (PMRF-192002-1174) of the Government of India. A.B is supported by the Core Research Grant (CRG/2023/005112) by the Department of Science and Technology Science and Engineering Research Board (India). R.K.S is supported by the Sabarmati Bridge Fellowship (MIS/IITGN-SBF/PHY/AB/2023-24/023) provided by Indian Institute of Technology Gandhinagar. We also thank the participants of the (virtual) workshop ``Testing Aspects of  General Relativity-II", ``Testing Aspects of  General Relativity-III" and ``New insights into particle physics from quantum information and gravitational waves" at Lethbridge University, Canada funded by McDonald Research Partnership-Building Workshop grant by McDonald Institute,  where some parts of the results are presented by A.C., for valuable discussions. A.B also like to thank the Department of Physics and Astronomy of University of Lethbridge, especially Saurya Das and  FISPAC Research Group, Department of Physics, University of Murcia, especially, Jose J. Fernández-Melgarejo for hospitality during the course of this work. A.B acknowledge associateship program of Indian Academy of Science, Bengaluru.

%%%%%%%%%%%%%%%%%%%%%%%%%%%%%%%%%%%%%%%%%%%%%%%%%%%%%%%%%%%%%%%%%%%
%%%%%%%%%%%%%%%%%%%%%%%%%%%%%%%%%%%%%%%%%%%%%%%%%%%%%%%%%%%%%%%%%%%
%\printbibliography

\end{document}